\documentclass[twocolumn]{aastex61}
\pdfoutput=1 
\usepackage{amsmath,amstext}
\usepackage[T1]{fontenc}
\usepackage{apjfonts} 
\usepackage[figure,figure*]{hypcap}
\usepackage{array,longtable}

\definecolor{gold}{HTML}{FFD700}
\definecolor{crimson}{HTML}{DC143C}
\definecolor{blue}{HTML}{008000}
\definecolor{violet}{HTML}{9400D3}

\let\oldtextbf\textbf
\renewcommand{\textbf}[1]{\textcolor{black}{#1}}

\shorttitle{Supergiants in globular clusters}
\shortauthors{Sz\'ecsi \& W\"unsch}

\begin{document}
	
    \title{Role of supergiants in the formation of globular clusters}
	
	\author{Dorottya Sz\'ecsi}
	\affiliation{Institute of Gravitational Wave Astronomy and School of Physics and Astronomy, University of Birmingham, Edgbaston, Birmingham B15 2TT, UK}
	\affiliation{Astronomical Institute of the Czech Academy of Sciences, Fri\v{c}ova 298, 25165 Ond\v{r}ejov, Czech Republic}
	\author{Richard W\"unsch}
	\affiliation{Astronomical Institute of the Czech Academy of Sciences, Bo\v{c}n\'{i} II 1401, 141 00 Prague 4, Czech Republic}
	
	\begin{abstract}
		Multiple stellar populations are observed in almost all globular-clusters, but the origin of this phenomenon is still debated. We investigate the role cool supergiants may have played. To do this, we combine two investigative methods: state-of-the-art massive stellar evolution and calculations of the hydrodynamic structure of the cluster-gas. This approach allows us to study how star-formation in young massive clusters depends on the energy- and mass-input of the first-generation of stars, while predicting the chemical composition of the second-generation. We find that the presence of massive (9$-$500~M$_{\odot}$) metal-poor supergiants in the young cluster leads to a star-formation episode \textbf{within the first 4~Myr of the cluster's lifetime, that is,} \textit{before} the first core-collapse supernovae explode \textbf{or the gas is expelled.} 
        The stellar winds accumulate in the cluster center, forming the second-generation there. Its composition is predicted to show variations in O~\&~Na~abundances, consistently with observations. The abundance of helium is, similarly to other scenarios involving massive stars, higher than what is referred from observations. Supposing dynamical removal of stars from the outskirts of the cluster, or applying a top-heavy initial-mass-function, we can predict a number ratio of the second-generation as high as 20$-$80\%. 
        The effect of metallicity is shown to be important, as the most luminous supergiants are only predicted at low-metallicity, thus limiting---but not excluding---the extent of a polluted second-generation at high-metallicity. These massive stars becoming black-holes suggests globular-clusters hosting gravitational-wave progenitors. Our scenario predicts a correlation between the mass of the cluster and the extent of the multiple population phenomenon.
	\end{abstract}
	
	
	
	\section{Introduction}
	
Young massive clusters (YMCs) are compact star forming regions with a radius of only a few parsecs \citep{PortegiesZwart:2010,Longmore:2014}. Since their projected lifetimes are consistent with those of old globular clusters \citep[GCs,][]{MaizApellaniz:2002}, they have been suggested to become GC-like objects eventually. In turn, old GCs, observed to populate the bulges and halos of many galaxies including our own, are hypothised to start out as massive clusters \citep{Brodie:2006,Andersen:2016}.
	
Both YMCs and GCs, as well as their suggested connection, are surrounded by observational puzzles. For example, why do we see multiple stellar populations in practically all GCs \citep[e.g.][]{Yong:2003,Gratton:2004,Harris:2010,DaCosta:2013,Bastian:2018}, and possibly in other clusters with ages up to 2~Gyr \citep[e.g.][]{Martocchia:2018}? Since one of the main indications that a cluster harbors multiple populations, is the anomalous ratios of light elements---\textbf{e.g. the observed ratio of sodium and oxygen, which} can only be synthesized at temperatures as high as 60$-$100~MK---it has long been suggested that a first generation of massive or intermediate mass stars is responsible for the formation of an anomalous second generation. But responsible in which sense? What are the conditions under which a second star formation episode can happen that feeds on the material ejected from the first generation? Or, to turn the question around, is the amount of material ejected from the first generation stars enough to produce the observed number of second generation stars? This latter puzzle is usually referred to as the `mass budget problem', since most scenarios suggested so far do struggle to answer~yes. 

It all may have something to do with metallicity, as we know that massive stellar evolution strongly depends on this \citep[e.g.][]{Meynet:2002,Yoon:2006,Brott:2011,Georgy:2013,Sanyal:2017,Vink:2018}. But how does the metallicity of the cluster influence the second generation star formation? 
In particular, how do the hydrodynamic structure of the cluster gas depend on the metallicity of the first stellar generation? How does the composition of the second generation depend on that?

Recently, \citet{Szecsi:2018} suggested that cool supergiants may play a role in the formation of the multiple populations in GCs. They investigated a scenario in which the second generation forms in a photoionization-confined shell around such cool supergiants, and speculated that the mass budget problem may be solved by only forming low-mass stars---\textbf{which, contrarily to some of the other scenarios that assumed the same \citep[e.g.][]{deMink:2009,DErcole:2010}, is better justified in these exotically shaped star forming regions \citep[cf. Sect.~4.7 of][]{Szecsi:2018}.} Nevertheless, they also suggested that even without such a shell, the wind of supergiant stars may play an important role in GCs that shall be more closely inspected---and this is the objective of present work. 
	
Here we investigate these questions by combining up-to-date theories of massive stellar evolution (those that predict cool supergiants) with calculations of the cluster's hydrodynamic structure. Earlier studies involving massive or intermedate mass stellar evolution were able to predict e.g. chemical pollution and element ratios \citep[e.g.][]{Karakas:2006, Decressin:2007, deMink:2009, Denissenkov:2014, Szecsi:2018}, but not able to say too much about the hydrodynamic behaviour of the gas in the cluster or, for that matter, under which conditions the formation of the second generation of stars happens. On the other hand, studies of the gas reinserted by massive stars within young clusters and its eventual accumulation leading to secondary star formation \citep[e.g.][]{Silich:2004, TenorioTagle:2005, Wunsch:2011, Wunsch:2017, Palous:2014, MartinezGonzalez:2016, Silich:2017} have hardly ever taken into account newly found peculiar aspects of stellar evolution. Combining these two research areas is therefore a viable and auspicious approach. 

To that end, we use a hydrodynamical, semi-analytic calculations of the cluster structure \citep[taken from][]{Wunsch:2017} which accounts for the winds of the first stellar generation as an input. We apply two different sets of single stellar evolutionary models for this first generation \citep[taken from][]{Brott:2011,Koehler:2015,Szecsi:2015}. They correspond to metallicities of the Large Magellanic Cloud (LMC) and of the low-metallicity dwarf galaxy I~Zwicky~18 (I~Zw~18). Thus we are able to investigate both the questions \textit{when?} and \textit{with which composition?} the second generation of stars may form in a young massive cluster. Additionally, we are able to study the process' dependence on metallicity, as well as the role that cool supergiants play in it.
	
This paper is organized as follows. The semi-analytic hydrodynamic code which determines the cluster structure, as well as the stellar evolutionary models, are described in Sect.~\ref{sec:Methods}. The synthetic population of stars that we created from the massive stellar models is discussed in Sect.~\ref{sec:Populations}. In Sect.~\ref{sec:Clusterwind} we perform calculations of the cluster's hydrodynamic structure applying the synthetic populations. We also discuss the conditions under which a second generation of stars may form. Sect.~\ref{sec:Composition} then investigates the chemical composition of this second generation. Sect.~\ref{sec:Massbudget} deals with the mass budget. Sect.~\ref{sec:discussion} discusses caveats and future directions, while Sect.~\ref{sec:conclusions} summarizes and concludes the work.
	

	\section{Methods}\label{sec:Methods}
	
	
	\subsection{Rapidly cooling shocked winds}\label{sec:windcalc}
	
Young massive clusters include large populations of massive stars concentrated in a rather small volume. Thus their stellar winds are expected to collide with each other and heat up to high ($\sim 10^6 - 10^7$\,K) temperatures. The overpressure of this hot gas then drives a star cluster wind \citep{Chevalier:1985}. If the cluster is massive and compact, and hence the density of the hot gas within it is high, the gas becomes thermally unstable, cools down to $\sim 10^4$\,K and forms dense clumps. The clumps are initially warm and ionized due to the radiation of nearby massive stars, but a fraction of them falls into the cluster centre due to the cluster gravity, where the gas accumulates until its column density is high enough to self-shield against the ionising radiation. Then, the gas cools further to lower temperatures and forms new stars. This scenario was explored extensively in a series of papers by \citet{Silich:2003,Silich:2004,TenorioTagle:2007,Wunsch:2008,Wunsch:2011,Palous:2013,Palous:2014} and others.
	
\citet{Wunsch:2017} studied this model of {\em rapidly cooling shocked stellar winds} by means of 3D hydrodynamic simulations including gravity (of both stars and gas), radiative cooling of the hot gas and EUV radiation of massive stars. They estimated a fraction of stellar winds that accumulates inside the cluster depending on various cluster parameters. 

They compared the results  of the 3D simulations to the outcome of a much simpler and much faster 1D semi-analytic code (see below) which is also able to estimate the mass of the second stellar generation. They found a good agreement. As for the first stellar generation, they relied on the predictions of the stellar synthesis code \textit{Starburst99} by \citet{Leitherer:1999}. They found that, using solar metallicity \textit{Starburst99} models of single stars up to M$_{\mathrm{top}}$~=~120~M$_{\odot}$ and a standard initial mass function, a substantially massive second generation of stars form only if the heating efficiency\footnote{Heating efficiency means the fraction of the mechanical energy of stellar winds that is transformed into thermal energy of the hot shocked gas inside the cluster.}, an observationally poorly constrained parameter, is very low. Here we apply the updated version of the semi-analytic code using stellar populations with different underlying physics (including different metallicities and M$_{\mathrm{top}}$). Below we shortly describe how the code works and what initial parameters we assume when running it.
	
In a gravitationally bound star cluster (SC), it can safely be supposed that properties of the stellar winds vary on a much longer timescale than the cluster wind crossing time. Therefore, one can search for a stationary solution of a set of spherically symmetric (1D) hydrodynamic equations to describe the \textit{star cluster wind} \citep{Chevalier:1985}. If the cluster is massive and compact enough, the hot gas is subject to radiative cooling due to its high density, and thus the set of stationary hydrodynamic equations to be solved should include the appropriate cooling term. A code to solve such a set with the assumption of spherical symmetry was developed first by \citet{Silich:2004}. Here we use a similar code described in \citep{Wunsch:2011,Wunsch:2017} updated with terms describing the effect of the star cluster gravity on the gas.

The set of stationary spherically symmetric hydrodynamic equations has a form
\begin{equation}
\frac{1}{r^2}\frac{d}{dr}(\rho u r^2) = q_m
\label{eq:basic_con}
\end{equation}

\begin{equation}
\rho u \frac{du}{dr} = - \frac{dP}{dr} - q_m u - \rho \frac{d\Psi_\star}{dr}
\label{eq:basic_mom}
\end{equation}

\begin{equation}
\frac{1}{r^2} \frac{d}{dr}\left[
\rho u r^2 \left( \frac{u^2}{2} + \frac{\gamma}{\gamma-1}\frac{P}{\rho} \right)
\right] = q_e - Q - \rho u \frac{d\Psi_\star}{dr}
\label{eq:basic_ener}
\end{equation}
where $\rho$, $u$ and $P$ are the wind density, velocity and pressure, respectively. The mass and energy input rate densities, $q_m$ and $q_e$ respectively, represent stellar winds approximated by a spatially smooth source described with a generalised Schuster distribution ($\sim [1+(r/R_c)^2]^{-\beta}$ for $r < R_\mathrm{SC}$) with parameters $R_c$, $R_\mathrm{SC}$ and $\beta$ being the core radius, cut-off radius and the slope of the distribution, respectively. The gravitational potential $\Psi_\star$ includes only a contribution from stars (i.e. the gas gravity is ignored) and it is assumed that the mass is distributed with the same generalised Schuster distribution as the wind sources and that the total mass is $M_\mathrm{SM}$. The cooling term has a form $Q = n_i n_e \Lambda(T, a_j)$ where $n_i = n_e = \rho/\mu_\mathrm{H}$ are the ion and electron number densities and $\Lambda(T,a_j)$ is a cooling function calculated by \citet{Schure:2009} with abundances of 15 species (H, He, C, N, O, Ne, Na, Mg, Al, Si, S, Ar, Ca, Fe, Ni), denoted $a_j$, calculated by the stellar population synthesis code (see Sect.~\ref{sec:popsyn}) from surface abundances of stellar evolution models (see Sect.~\ref{sec:stellar}).

The solution of Eqs.~(\ref{eq:basic_con})--(\ref{eq:basic_ener}) exists only in a subset of the parameter space. It is possible to define a critical luminosity $L_\mathrm{crit}$ so that the solution exists only if the total mechanical luminosity of all stellar winds is smaller than the critical value, $L_\mathrm{SC} < L_\mathrm{crit}$. The semi-analytic code determines the critical luminosity iteratively by trying to solve Eqs. (\ref{eq:basic_con})--(\ref{eq:basic_ener}) for a given set of parameters: $R_c$, $R_\mathrm{SC}$, $\beta$ and properties of the wind of a stellar population with a given mass, age and chemical composition.  It has been shown by \citet{Wunsch:2017} that $L_\mathrm{crit}$ can also be used to estimate the rate of clump formation: the mass of clumps formed over time is the difference between $\dot{\mathrm{M}}_\mathrm{SC}$ and $\dot{\mathrm{M}}_\mathrm{crit}$, the former being the mass deposition rate of the cluster, and the latter the corresponding quantity but taking L$_\mathrm{SC}$~$=$~L$_\mathrm{crit}$. The mass accumulation rate, $\dot{M}_\mathrm{acc}$ is then determined by taking into account only clumps which are formed with initial velocity (the same as the cluster wind velocity $u$) smaller than the escape velocity $u_\mathrm{esc} \equiv \sqrt{2\Psi_\star}$

As the semi-analytic code models the star cluster using a smooth distribution of mass and energy sources, the calculations cannot represent discrete events such as supernova explosions nor, therefore, the effect of the dust they produce. We discuss why and when it is justified to omit supernovae in our calculations in Sect.~\ref{sec:supernovae}. 	

In this work, we discuss two cluster models differing in the initial chemical composition of massive stars (cf.~Sect.~\ref{sec:stellar}). We assume that all first generation stars are formed abruptly at $t = 0$ and we follow the cluster evolution for $10$\,Myr. The total mass of the first generation stars is  $M_{\mathrm{SC}}$~=~10$^{7}$~M$_{\odot}$ for both models, the stars are assumed to form with the standard Initial Mass Function (IMF, \citealt{Kroupa:2001}), and they are represented by the stellar models described in Sect.~\ref{sec:stellar}. The second generation stars are represented only as the accumulated mass ($M_\mathrm{acc}$), they do not contribute to our calculations by e.g. their stellar winds. The stellar density profile of the cluster is given by the generalised Schuster distribution with $R_c = 1$\,pc, $R_\mathrm{SC} = 3$\,pc, and $\beta = 1.5$. 
As opposed to \citet{Wunsch:2017}, here we assume that the all mechanical energy of stellar winds is converted to the thermal energy of the hot gas (i.e. the heating efficiency is unity), and that the mass loading is zero. On the other hand, we use different upper mass limit, M$_{\mathrm{top}}$, of our first generation stellar population, as explained in Sect.~\ref{sec:popsyn}. Additionally, we carry out a parameter space study (details given in Sect.~\ref{sec:paramspace}), where we vary the initial cluster mass, $M_\mathrm{SC}$, and the index of the IMF for stars more massive than $1$\,M$_\odot$. 
	
	\subsection{Stellar evolutionary models}\label{sec:stellar}

\begin{figure*}
		\centering
		\includegraphics[width=.95\columnwidth]{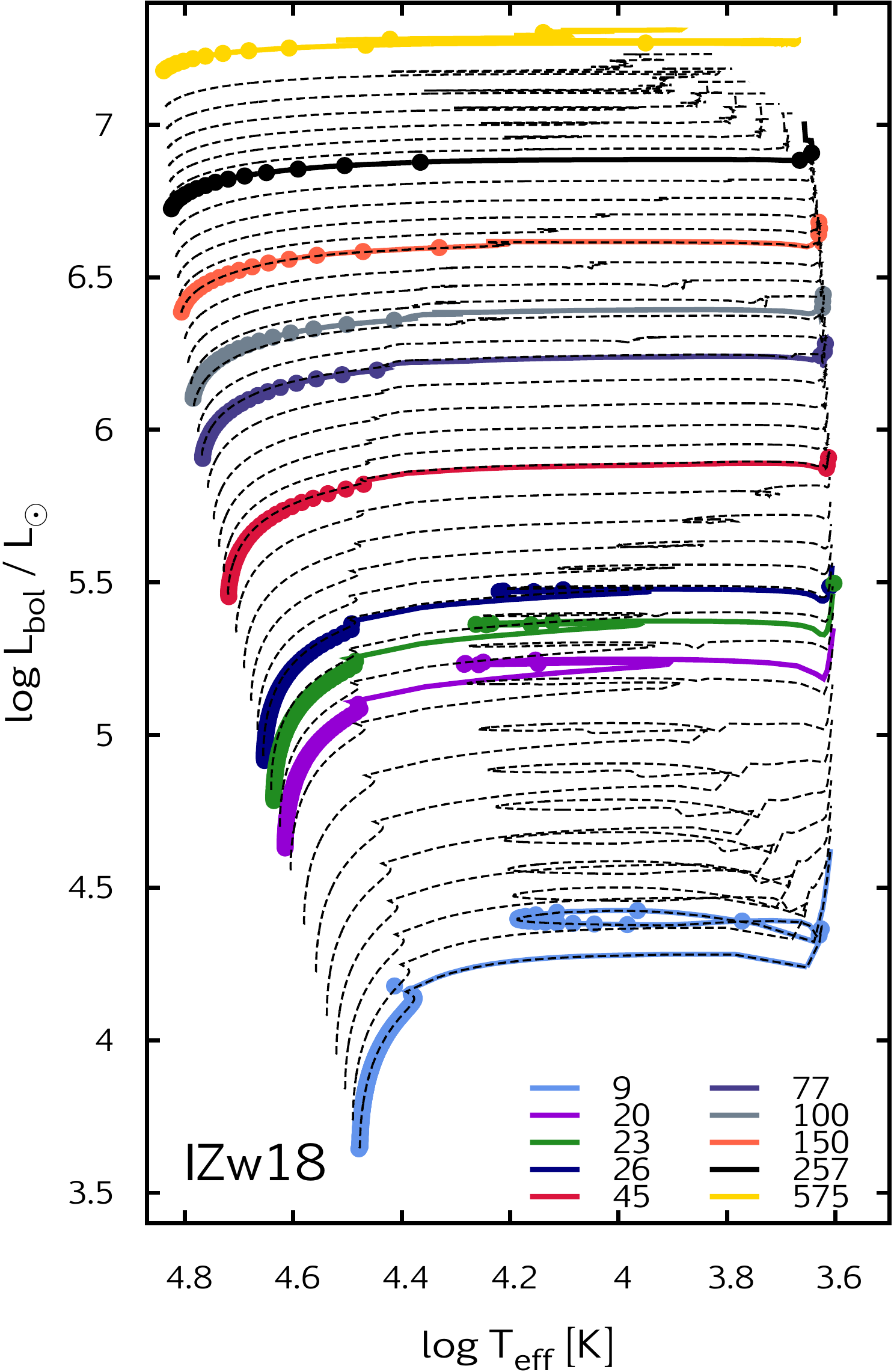} \hspace{40pt}
		\includegraphics[width=.95\columnwidth]{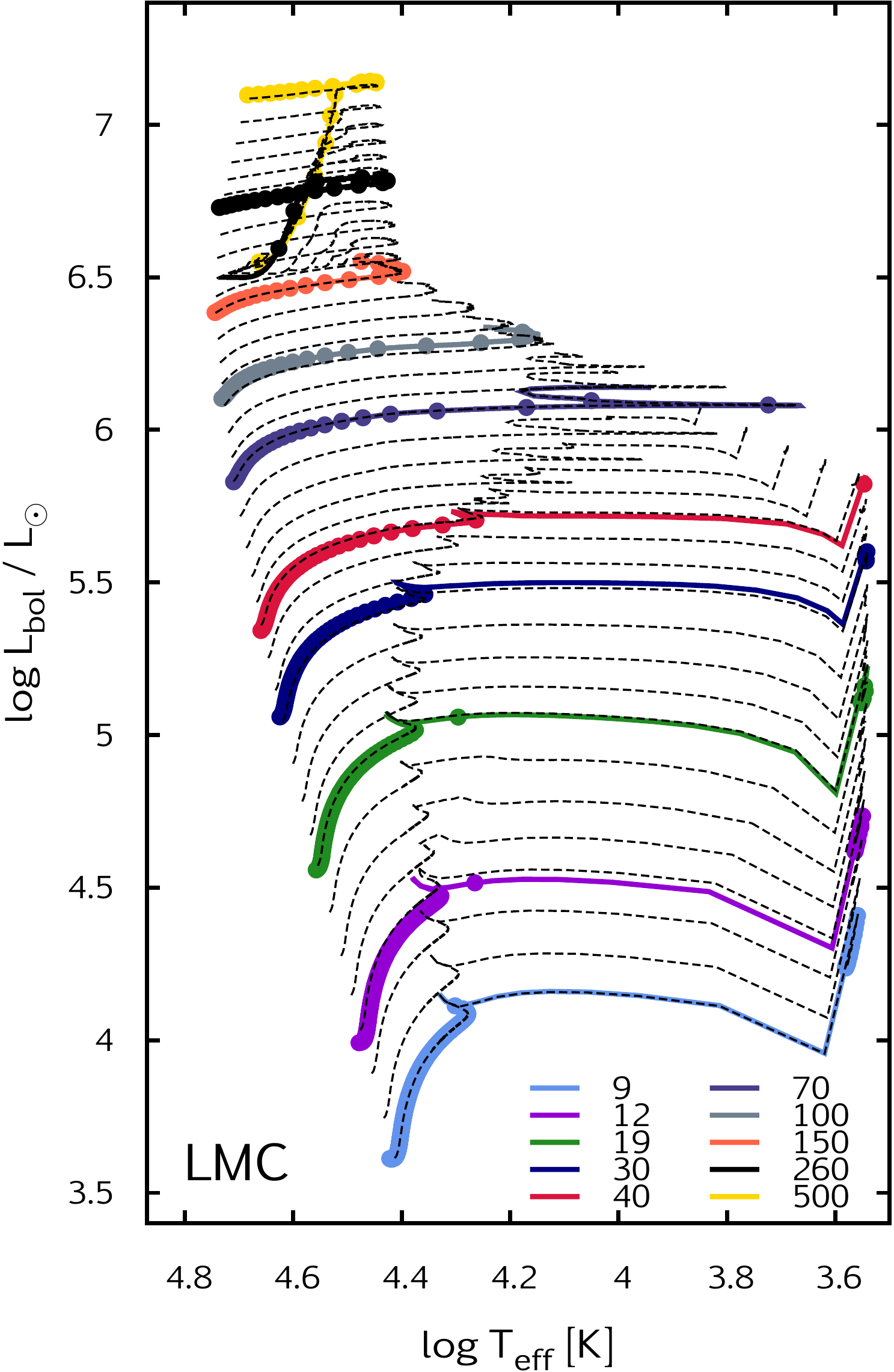}
		\caption{Hertzsprung--Russell diagrams of the two sets of models. Left: low-metallicity models, rigth: high-metallicity models. Initial masses are color coded and indicated by the key legends (units in M$_{\odot}$). Black dashed lines mark interpolated tracks between the colored models; interpolation is performed between 0.1--500~M$_{\odot}$, resulting $\sim$ 5000~tracks in this mass range.}\label{fig:HRDs}
	\end{figure*}
	
	To account for the first generation of massive stars, we apply two sets of models, both computed with the BEC code \citep[see e.g.][and references therein]{HegerI:2000,Yoon:2012,Szecsi:2015}. The models with initial composition of the LMC were created by \citet{Brott:2011a} and \citet{Koehler:2015}, representing a subsolar-metallicity environment with $\sim$0.4~Z$_{\odot}$ (i.e. [Fe/H]~$\sim$~$-$0.4). Those with initial composition of the dwarf galaxy I~Zw~18 
    were created by \citet{Szecsi:2015} and \citet{Szecsi:2018}, representing a low-metallicity environment with $\sim$0.02~Z$_{\odot}$ (i.e. [Fe/H]~$\sim$~$-$1.7). 	
    
    The low-metallicity models between M$_{\mathrm{ini}}$~$=$~10--300~M$_{\odot}$ have all been followed until the end of core-helium-burning \citep[completing on the work of][who only followed them until the end of core-hydrogen-burning]{Szecsi:2015}. From this point on, we mainly refer to the composition of the LMC as high-metallicity and to that of I~Zw~18 as low-metallicity.

	The most massive models ($>$70~M$_{\odot}$ in the high-metallicity set and $>$300~M$_{\odot}$ in the low-metallicity set) have been, however, only computed until core-hydrogen-exhaustion (i.e. terminal-age main-sequence). Therefore, we extrapolate for how much mass they would lose during their remaining evolution if the mass-loss rate was the same as that at the end of the computation. This is clearly a simplistic approach that brings some additional uncertainty into our predictions.
	
	As for rotation, all the models have zero or slow rotation (i.e. 0~or 100~km~s$^{-1}$ initially). They evolve with a distinct core-envelope structure towards lower surface temperatures. It shall be a future task to add models that have more extreme rotation rates (and, for example, evolve chemically-homogeneously). Also, effects of binarity are omitted at this point as we only apply single stellar models. 

The wind velocity, $v_\mathrm{wind}$, of any given stellar model is calculated from the escape velocity from the stellar surface as v$_{\mathrm{wind}}$~$=$~1.3$\cdot$v$_{\mathrm{esc}}$ and v$_{\mathrm{wind}}$~$=$~2.6$\cdot$v$_{\mathrm{esc}}$ for models below and above a surface temperature of 21~kK, respectively, following the theory of line driven winds \citep{Lamers:1999}. Additionally, following e.g. \citet{Leitherer:1992}, the wind velocity is corrected for the metallicity of the wind material, $Z$, by multiplying it by a factor $(Z/Z_\odot)^{0.13}$ .  Since supergiants' winds are not expected to be line driven, we checked that the outcome of our calculations is not, in fact, sensitive to the actual values of supergiant wind velocity we use, as long as they are below 80~km~s$^{-1}$, which they indeed are.


	\section{Stellar populations}\label{sec:Populations}

	\subsection{Comparing the two sets of models}\label{sec:grids}
	
	Fig.~\ref{fig:HRDs} shows the Hertzsprung--Russell diagrams of the two sets of models. The most important difference between the high-metallicity models (LMC) and the low-metallicity models (I~Zw~18) is the presence of very massive ($\gtrsim$100~M$_{\odot}$) cool supergiants in the latter case, populating the top-right corner of the diagram. The reason the very massive, high-metallicity models do not evolve to the supergiant branch is that their mass-loss rate during the main-sequence is high enough to remove almost the whole envelope, turning them into hot stars such as Luminous Blue Variables or Wolf--Rayet stars. As we show in Sect.~\ref{sec:Macc-lowZ}, the presence of very massive supergiants at low metallicity is the key to form a second generation of stars early enough so that they show chemical abundances attributed to a subsequent population.
	
	Other differences between the high- and low-metallicity sets of models are: 
	
	\begin{itemize}\setlength\itemsep{-2pt}
		\item both the zero-age main-sequence and the terminal-age main-sequence lie at lower surface temperatures when the metallicity gets higher;
		\item blue supergiants are found in the low-metallicity models with 9--30~M$_{\odot}$ during core-helium-burning (i.e. blue loop); 
        \item red supergiants with 9--45~M$_{\odot}$ are found amongst the high-metallicity models, these are also core-helium-burning objects;
		\item the presence of Luminous Blue Variables at M$_{\mathrm{ini}}$~$\sim$~70-100~M$_{\odot}$ in the high-metallicity models; Luminous Blue Variables are found in the low-metallicity models only at masses above $\gtrsim$~300~M$_{\odot}$;
		\item the presence of Wolf--Rayet stars at M$_{\mathrm{ini}}$~$\gtrsim$~100~M$_{\odot}$ in the high-metallicity models; no Wolf--Rayet stars are found in the low-metallicity models.
	\end{itemize}
	
Since the very massive supergiants at low-metallicity are the key objects responsible for the multiple population phenomenon in our model, it is important to discuss them here a bit further. 

\subsection{Cool supergiants}

Low-metallicity models with M$_{\mathrm{ini}}$~$\gtrsim$~80~M$_{\odot}$ evolve to the supergiant branch even during their core-hydrogen-burning phase due to envelope inflation. This phenomenon has been investigated by \citet{Sanyal:2015,Sanyal:2017} who found that the reason these stars inflate their envelopes and thus expand is their proximity to the Eddington-limit. 

What is extremely intriguing in the context of the chemical composition and the multiple populations in star clusters, is that these massive stars becoming cool supergiants during their core-hydrogen burning phase means they have a convective envelope. Convection mixes the material between the core---where the CNO-cycle operates, together with side reactions that can synthesise Na and Al at the expense of O and Mg---and the surface. The surface layers are then removed by the stellar wind, thus polluting the interstellar gas with nuclear ashes that have undergone hot-hydrogen-burning. Interestingly, the convective core does not reach down again into the burning regions during core-helium burning (since these layers are much deeper inside than those of core-hydrogen burning), thus avoiding the ejection of helium-burning products. This makes these supergiants quite ideal to be suggested as potential pollution sources in GCs.

If these stars exist in nature, is a question for future investigations. Envelope inflation has been recently studied by several authors \citep{Grafener:2012,Grassitelli:2015a,Grassitelli:2015b,Sanyal:2015,Sanyal:2017}. In particular, \citet{Moriya:2015} suggested that such supergiants may be responsible for some supernovae of the superluminous type. 	
An additional caveat of using simulations of supergiants is that their mass loss rates are, despite great efforts to constrain them, still quite uncertain. For a comprehensive discussion on the subject, we refer to the book of \citet{Levesque:2017} as well as to the relevant literature on our stellar models (Sect.~5 in \citealt{Szecsi:2015} and Sect.~2.1 in \citealt{Szecsi:2018}).
	
	\subsection{Population synthesis}\label{sec:popsyn}

To create a synthetic population, we suppose that all the first generation stars of the cluster were formed during a single starburst episode, almost instantaneously. We assume a standard piece-wise power-law IMF with three intervals ($0.01 - 0.08$, $0.08 - 0.5$ and above $0.5$\,M$_\odot$) with indeces $-0.3$, $-1.3$ and $-2.3$, respectively, suggested by \citet{Kroupa:2001}. In Sect.~\ref{sec:paramspace} we explore how the results change if a top-heavy IMF is used, by introducing an additional fourth interval for masses above $1$\,M$_\odot$ with index, $\alpha_4$, varying between $-2.3$ and $-1.1$.

	Since both sets of stellar evolutionary models only contain ten models between 9--500~M$_{\odot}$, we need to interpolate between these tracks to get two, smoothly changing grids (cf. dashed lines in Fig.~\ref{fig:HRDs}). We split the whole range of stellar masses (0.1--500~M$_{\odot}$)\footnote{We use a I~Zw~18 model with 575~M$_{\odot}$ in the interpolation, but still take the upper limit M$_{\mathrm{top}}$~$=$~500~M$_{\odot}$ for the synthetic population, to be consistent.} equidistantly in the logarithmic scale into 5000 intervals.
    For intervals above $9$\,M$_\odot$ (there is $\sim$1800 of them) we apply the interpolated tracks, while stars less massive than that are taken into account only as mass holders. When interpolating between the stellar models, we use logarithmic scale to interpolate linearly between age, mass, mass-loss rate, surface temperature, radius and surface abundances; while bolometric luminosity, escape velocity and wind energy rate are calculated as follows: 
	$L_{bol} = \sigma_{SB} \cdot T_{eff}^4 \cdot 4\pi \cdot R^2$;
	$v_{esc}^2 = 2 G\cdot M/R$ and
	$L = 0.5 \cdot \dot{M} \cdot v_{esc}^2$.
The tracks of this smooth grid are weighted with the IMF and interpolated for a selected age. Then sums of various quantities, such as mass and energy deposition rates, and mass weighted mean abundances, are calculated. This calculation is carried out for $500$ points distributed uniformly throughout the followed period of the cluster evolution ($10$\,Myr). The resulting time evolutions of these quantities is then used as the input for the semi-analytic code (described in Sect.~\ref{sec:windcalc}) which computes the structure of the star cluster wind and eventually the rate of the mass accumulation and its chemical composition for each point in time. 
	
	We sampled the stellar evolution models so that major steps occur more or less at the same evolutionary stage; in particular, since in our cluster calculations the most important property is the mass-loss rate, we made sure that the interpolated mass-loss rate behaves well around the bi-stability jump \citep[i.e. at T$_{eff}$~$\sim$~21~kK where the mass loss rates change abruptly due to an increase in the line acceleration of Fe~III below the sonic point of the stellar wind, cf.][]{Vink:1999,Vink:2000}. 
	
	\begin{figure*}
		\centering
		\includegraphics[height=.99\columnwidth,angle=270]{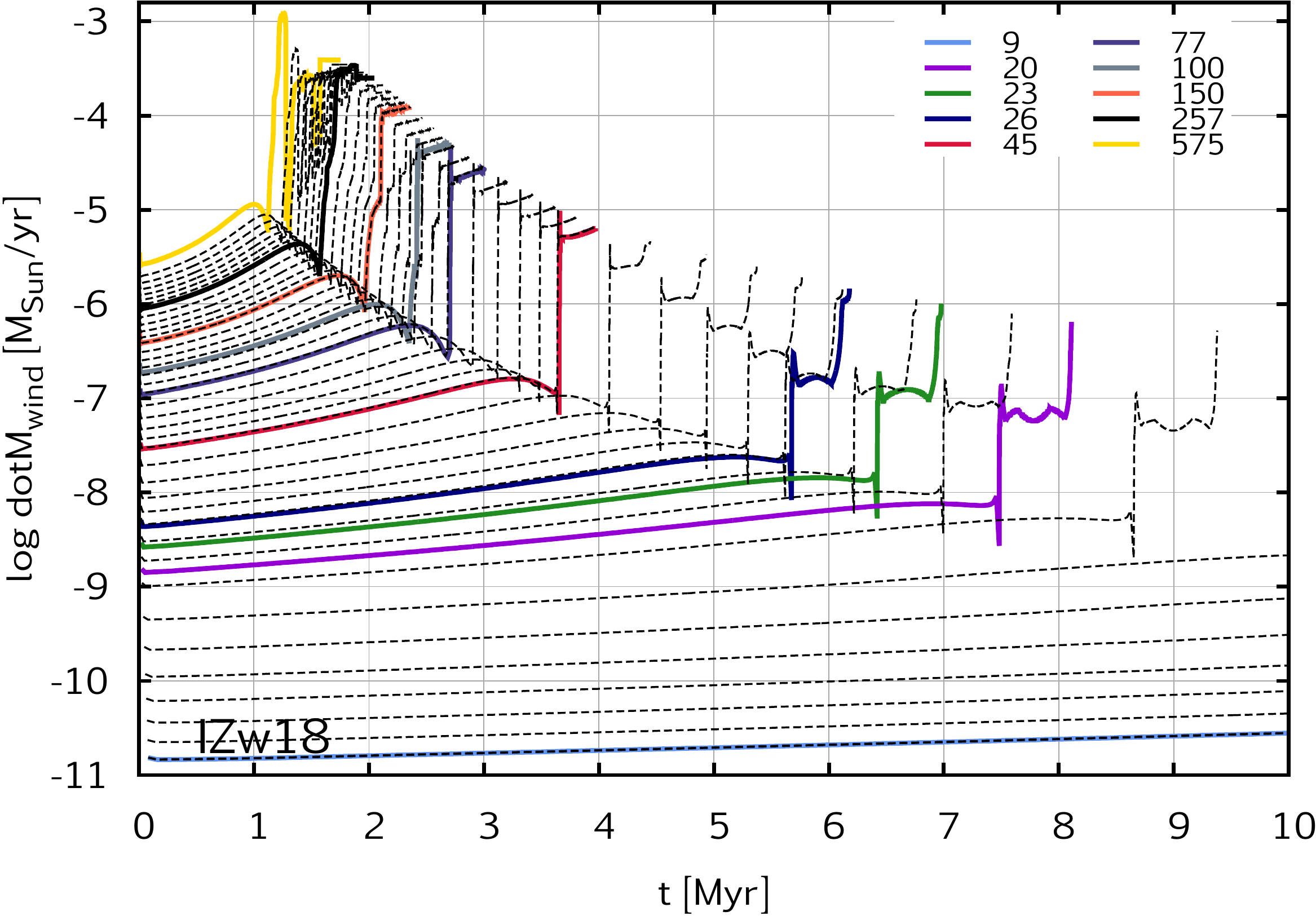} \hspace{10pt}
		\includegraphics[height=.99\columnwidth,angle=270]{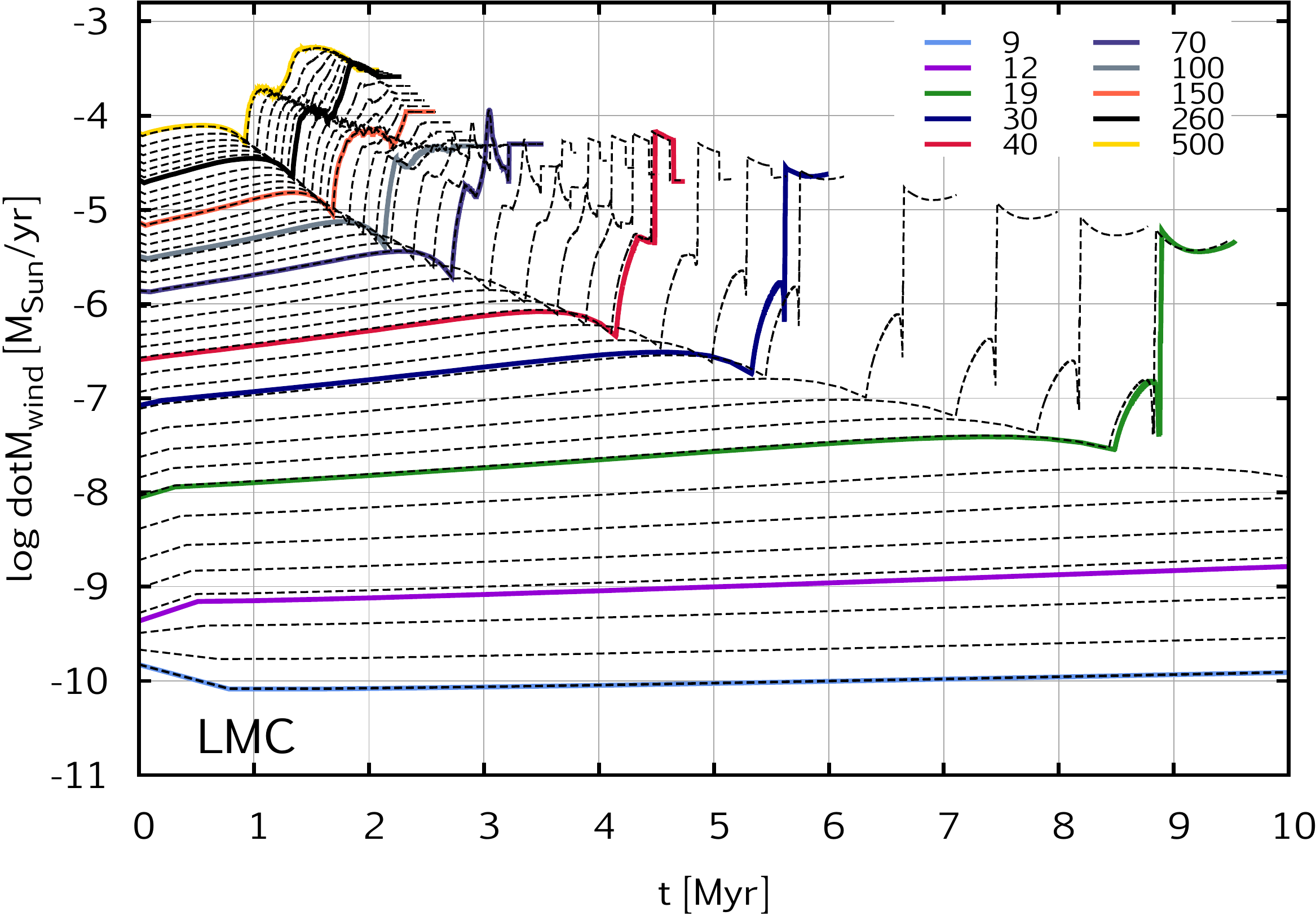} \\
		\includegraphics[height=.99\columnwidth,angle=270]{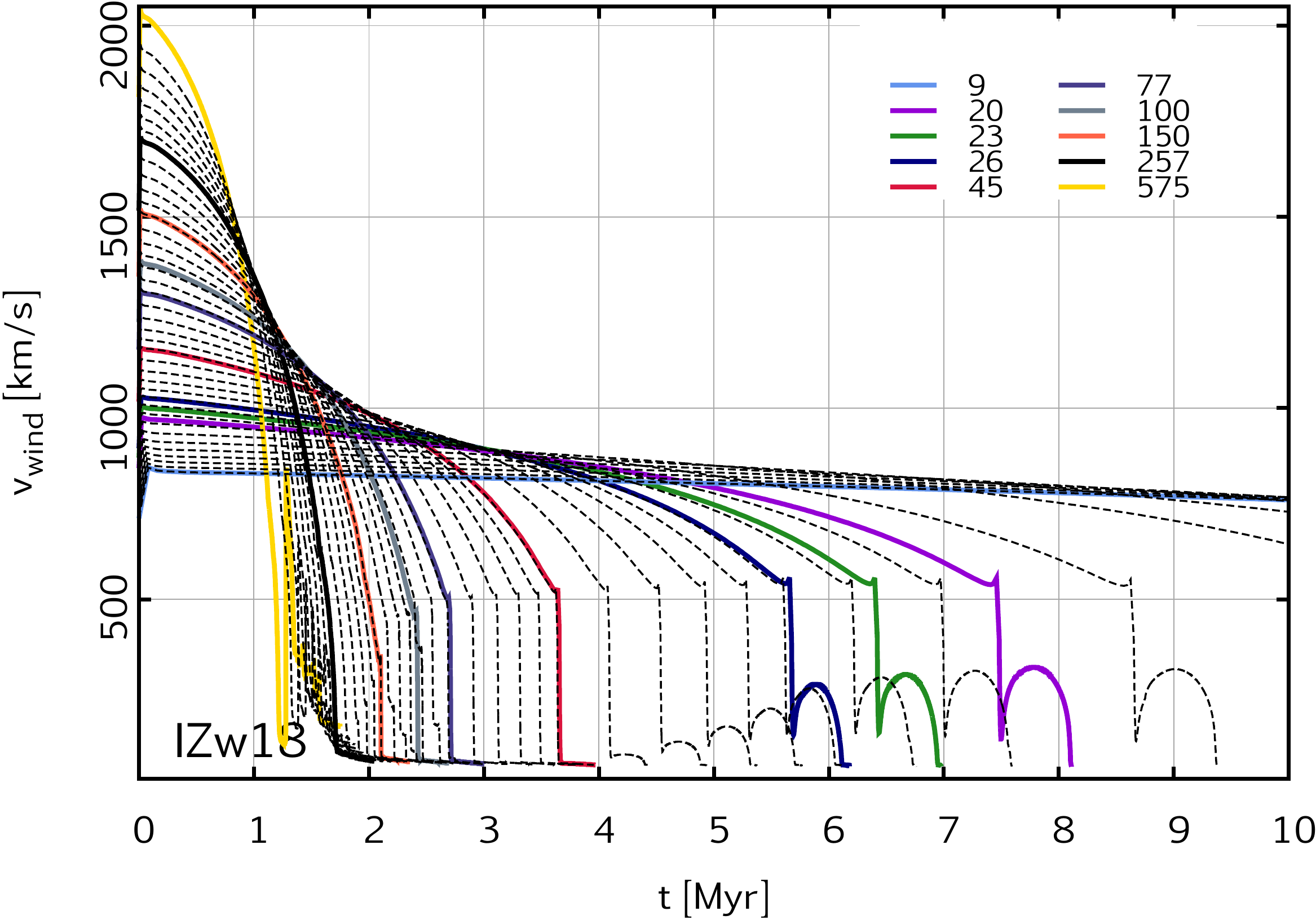} \hspace{10pt}
		\includegraphics[height=.99\columnwidth,angle=270]{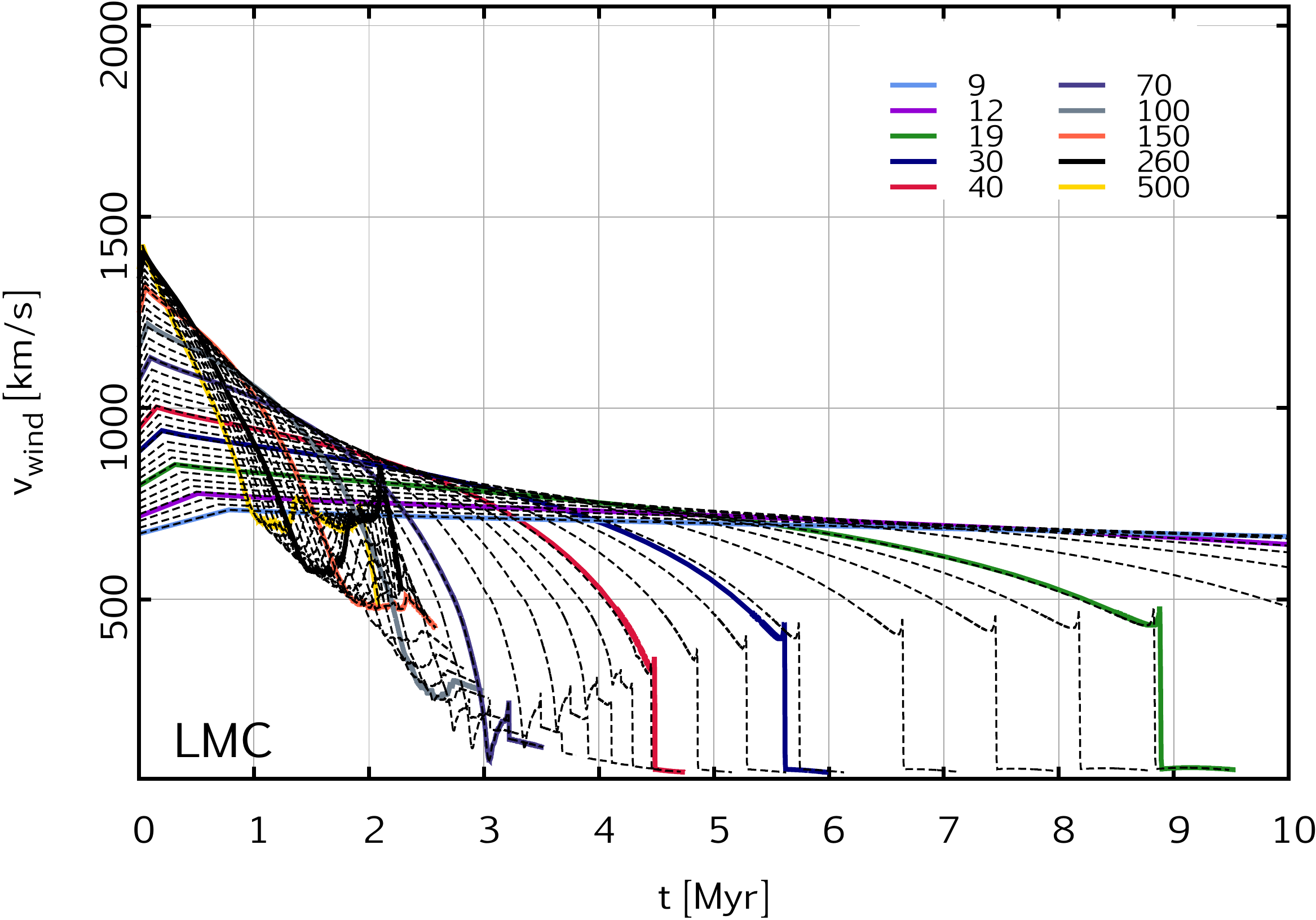}\\	
		\includegraphics[height=.99\columnwidth,angle=270]{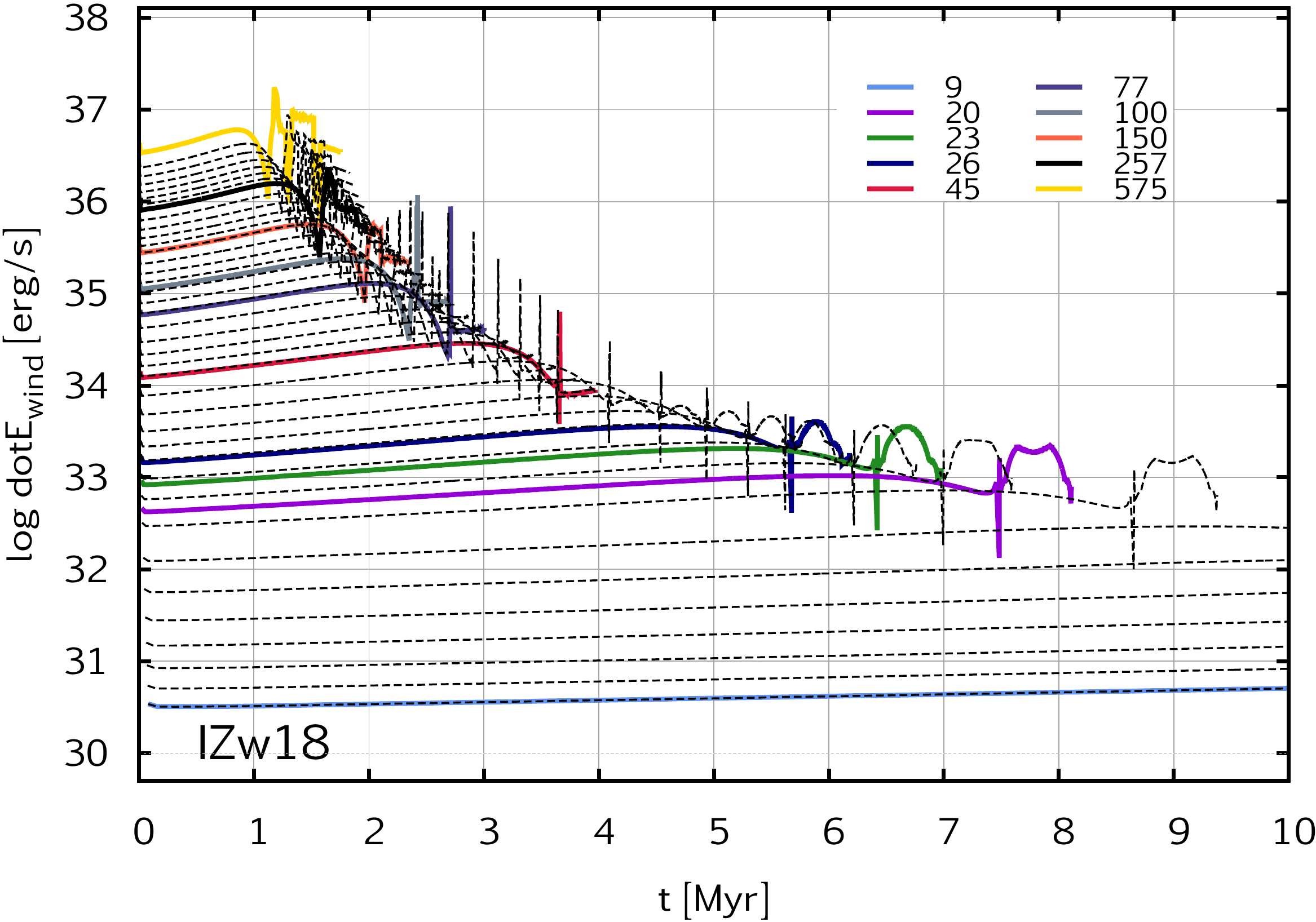} \hspace{10pt}
		\includegraphics[height=.99\columnwidth,angle=270]{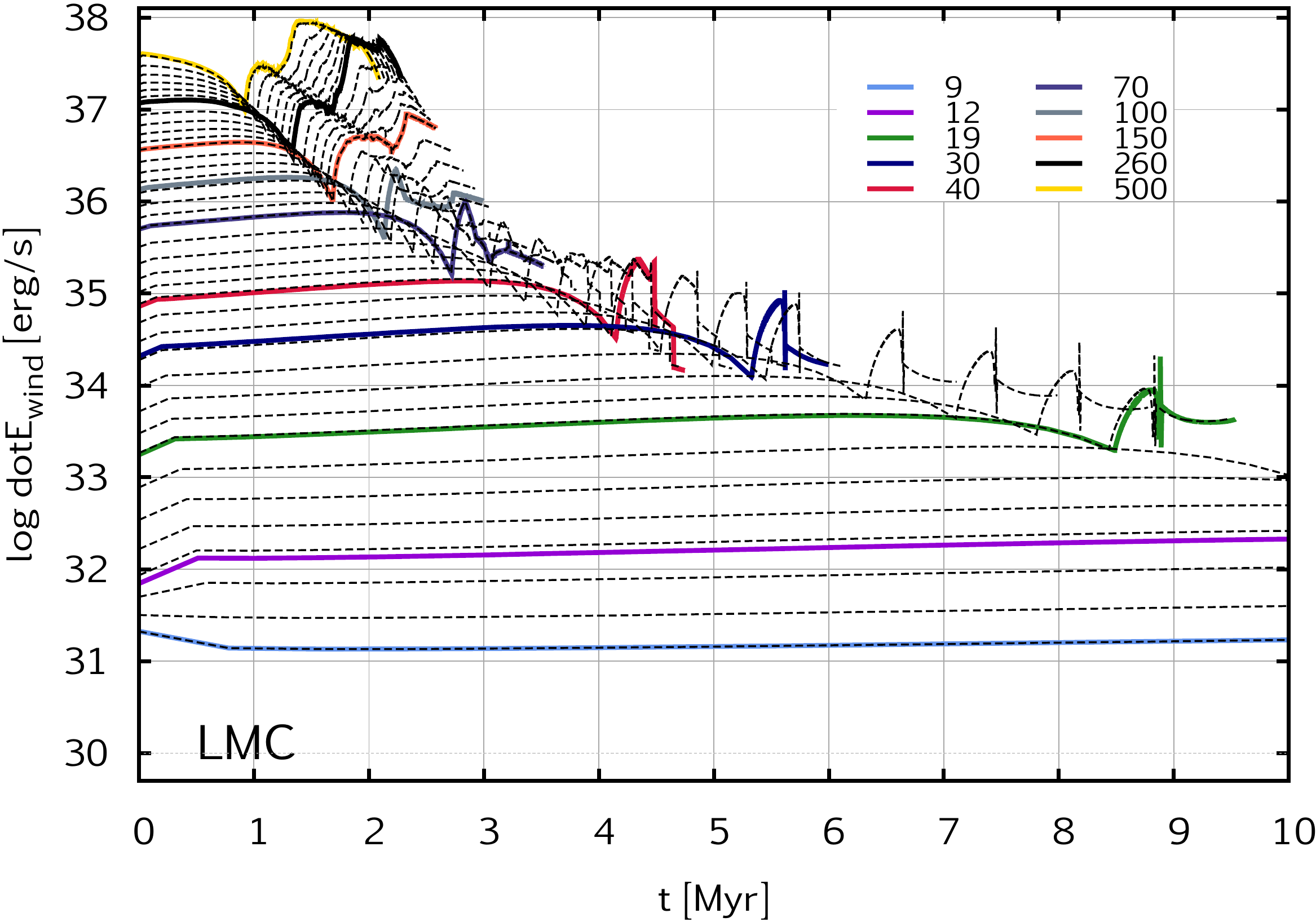}
		\caption{Time evolution of our low- and high-metallicity stellar populations (left and right panels, respectively) during the first 10~Myr of the clusters' life. Colored lines are taken from \citet{Szecsi:2015} and \citet{Szecsi:2018}, while black dashed lines are interpolated tracks. 
		Top figures show the mass loss rates in the models; middle the terminal velocity of their stellar winds; bottom the energy flux these winds insert into the cluster gas. 
			}\label{fig:int}
	\end{figure*}
	
As opposed to earlier works on the same subject \citep[such as][]{Wunsch:2017} who used M$_{\mathrm{top}}$~$=$~120~M$_{\odot}$, here we include very massive stars up to M$_{\mathrm{top}}$~$=$~500~M$_{\odot}$. This choice is motivated by the findings that 
low-metallicity stellar models between 150--500~M$_{\odot}$ display significant variations in their surface abundances of light elements \citep{Szecsi:2018}. 
There are observational implications for the existence of stars up to 315~M$_{\odot}$ in the LMC \citep{Crowther:2010,Crowther:2016}. At low-metallicity, radiation driven winds are less effective, so ideally even more massive stars than that may form. 

	\subsection{Wind properties of the populations}
	
Figure~\ref{fig:int} presents the properties of the stellar wind such as its mass loss rate, mechanical luminosity, as well as its velocity for all model stars in the population. The mass loss rate typically becomes higher with stellar mass. While slowly increasing in the first half of the stars' life, it then experiences a local minimum and then a sudden jump. This jump is attributed to certain changes in the wind structure at T$_{\mathrm{eff}}$~$\sim$~21~kK leading to an increased mass loss. \citep[It is called the bi-stability jump, and concerns the fact that additional iron line transitions become effective in driving the wind under this temperature; see][]{Vink:1999}. Soon after this, the stars reach their post-main-sequence phase, during which the mass loss rates are 1-2 orders of magnitude higher than during the main-sequence phase. The models' evolution is computed until core-hydrogen exhaustion, but subsequent evolutionary phases are very short ($<$~1\%) compared to the total lifetimes, so they can be safely omitted from considerations of stellar wind mass loss. 
	
Models with low-metallicity have typically lower mass loss rates than their high-metallicity counterparts of the same mass. The reason is that the mass loss rate is not only the function of stellar mass, but also of metallicity (and, at some extent, of other stellar paratmeters such as radius and surface composition). The mass loss rate's dependence of metallicity is prescribed as $\dot{M}\sim Z^{0.86}$ in both set of models at every evolutionary phase, while its dependence on the actual stellar mass follows $\dot{M}\sim M^{1.13}$. 

There is one evolutionary phase when the mass loss experienced by low-metallicity models is higher than that experienced by the high-metallicity ones: in the case of the most massive models' late evolution (i.e. above 100~M$_{\odot}$). Here the mass loss of the high-metallicity models follow a prescription typical for Wolf--Rayet stars, while that of the low-metallicity ones follow a prescription typical for red supergiants. While the latter does not predict significanty higher mass losses than the former, we have to take into account another effect also playing a role. Namely, that since the mass loss rates during the first half of the main sequence are lower in the case of low-metallicity models, these stars are typically more massive during their later phases than those predicted by high-metallicity models of the same initial mass, and therefore their mass loss rates are now higher. 
The jumpy mass loss rates visible for example during the late phases of the high-metallicity model with 70~M$_{\odot}$ or the low-metallicity model with 575~M$_{\odot}$, are due to these models being associated with an LBV phase.
Models below 30~M$_{\odot}$ at low-metallicity experience a blue loop during their core-helium-burning phase. The mass loss attributed to this blue supergiant phase is typically lower than what is expected for a red supergiant, leading to a plateau in these models' post-main-sequence mass loss rates. 
	
The behaviour of wind velocity of the models in Fig.~\ref{fig:int} can be understood as follows. Initially, the low-metallicity models are hotter than the high-metallicity ones due to their surface opacity being lower. Therefore, these models are typically smaller in radial size. As wind velocity is computed from the escape velocity, it is expected that their wind velocity is higher than their high-metallicity counterparts'. As the evolution progresses however, the low-metallicity models evolve towards lower surface temperatures and larger radii, thus their wind velocity drops. The same is happening with the high-metallicity models. One crucial difference is that while the low-metallicity models become supergiants, the high-metallicity ones, at least those above 70~M$_{\odot}$, become LBVs or WR~stars. These models' wind velocity is high. As for masses below 70~M$_{\odot}$, both grids predict supergiants---the low-metallicity grid blue supergiants below 40~M$_{\odot}$ with a loop in the wind velocity. 
	
Energy flux of the wind is closely related to both the mass loss and the wind velocity, as it is computed by these two as ${\mathrm{L}_{\mathrm{wind}}}$~$=$~1/2~$\dot{\mathrm{M}}$~v$_{\mathrm{wind}}^2$. From this we can calculate the mechanical luminosity inserted into the cluster by the first generation of stars, by properly considering every individual stellar model's contribution to the population (i.e. weighting with the IMF). Thus we arrive to the value L$_{\mathrm{SC}}$ presented in Fig.~\ref{fig:lum}. 
	
	\subsection{Supernova explosions}\label{sec:supernovae}
	
Massive stars end their lives in various ways, depending on mass and metallicity. If supernova explosions happen, they may contribute to the cluster's subsequent evolution significantly. Core-collapse supernova explosions in particular may increase the cluster's iron content. For this to happen however, a mechanism is needed that traps the (possibly very energetic) supernova ejecta inside the cluster's potential well. Such mechanism was suggested e.g. by \citet{TenorioTagle:2013} and further studied by \citet{MartinezGonzalez:2018}, however, its discussion is out of the scope of this work.
	
Our semi-analytic model does not include this effect. Therefore, we need to discuss when and what kind of supernova explosions we expect from our massive stellar models (if any) so that we can carefully evaluate the validity of our semi-analytical approach. \textbf{The majority of globular clusters show no variation in their iron content, so to account for them with our model, we need to play extra attention to the time periods when supernovae's contribution enter the picture. Some globular clusters such as e.g. $\omega$~Cen and M~54 are peculiar in this regard, displaying variations in iron and even in their total sum of C, N and O content. Still, their formation may have happened fundamentally differently from the average cluster \citep[$\omega$~Cen in particular was suggested to start out as a dwarf galaxy, cf.][]{Schiavon:2017}. Therefore we only account for the majority of clusters here, those that do not exhibit star-to-star variations in iron abundance.}
	
Connecting stellar models to supernova types is somewhat uncertain. It depends on both the model in question, and the assumptions about the nature of the explosion. Here we simply rely on the work of \citet{Heger:2003} and establish that, in the case of our low-metallicity population, the first supernova that may pollute the cluster with iron (which is a 40~M$_{\odot}$ star), explodes at a cluster age of 4.5~Myr. The reasons are the following. 
	
In the low-metallicity population, massive stars up to an initial mass of 25~M$_{\odot}$ are expected to explode as core-collapse supernovae of type~II~P. The total lifetime of a 25~M$_{\odot}$ star is 7~Myr; those at lower masses live even longer. We expect them to form neutron stars as remnants. As for stars between an initial mass of 25$-$40~M$_{\odot}$, they are expected to explode as weak II~P supernovae and form black holes as remnants. The total lifetime of a 40~M$_{\odot}$ star is about 4.5~Myr. 
	
Above this mass but below 140~M$_{\odot}$, it is expected that the metal-poor models do not explode but fall into a black hole directly due to their immense self-gravity at the moment of iron-core collapse. An initially 140~M$_{\odot}$ model has a total lifetime of 2.4~Myr. This means that stars with lifetimes between 2.4 and 4.5~Myr do not explode as supernovae; athough they contribute to the (stellar-mass) black hole content of their clusters. The same fate awaits those stars that have an initial mass above 260~M$_{\odot}$---that means many of our metal-poor models form black holes without an explosion. 
	
Between 140$-$260~M$_{\odot}$ (i.e. total lifetimes of 1.9$-$2.4~Myr), the models in the low-metallicity set are again predicted to explode. This time though, it is not due to iron-core collapse but another effect: pair-creation\footnote{It is not relevant to the present discussion, but those low-metallicity stars above 260~M$_{\odot}$ that fall into a black hole directly, do so also due to pair-creation induced instability and not an iron-core collapse.}. During their oxygen-burning phase, the creation of electron$-$positron pairs disturbs their hydrostatic stability, and makes them explode without leaving a remnant. Such a pair-instability supernova does not pollute the cluster with iron, as nuclear fusion has not yet produced an iron-core. The core that explodes contains mainly carbon and oxygen, which undergo explosive burning during the supernova event, producing a unique nucleosynthetic signature \citep{Burbidge:1957,Langer:1991,Heger:2003,Langer:2007,Kozyreva:2014}.
	
A pair-instability supernova is expected only for a small number of all massive stars in our low-metallicity population (for a 10$^7$~M$_{\odot}$ cluster with the standard IMF, we expect about 1 such supernova in every 270~years). 
For our present purposes, we do not investigate how these supernovae may contribute to our cluster's structure or chemical composition, but consider all stars above an initial mass of 40~M$_{\odot}$ (that is, an age of 4.5~Myr) not to pollute the cluster with iron. We also do not investigate the effect of pulsational pair-instability \citep{Woosley:2007,Moriya:2015}, pointing out that this process may also play some, yet to be investigated,  role in polluting the cluster.
	
As for the high-metallicity set of models, the situation is quite different. Here we also expect type~II~P supernovae from models between 10$-$25~M$_{\odot}$ initial mass, with stars between 25$-$40~M$_{\odot}$ ending up as a type~II~L/b supernovae. But above that limit, instead of falling into black hole due to self-gravity or undergoing pair-creation induced instability, stars are expected to explode as supernovae of type~I~b/c---that is, due to iron-core collapse. The reason for the high-metallicity models exploding as core-collapse supernovae instead of some more exotic scenario, is that they lose so much mass during their lifetimes that their final mass is in the range where neither self-gravity is strong for direct black hole formation, nor is pair-creation playing a role. So, all our high-metallicity stellar models undergo a core-collapse induced supernova explosion that may, if the ejecta is trapped in the cluster, pollute the gas with iron. The first supernova, that of our most massive model (initial mass 500~M$_{\odot}$ but final mass only 34~$M_{\odot}$) explodes at the age of 2.1~Myr.
	
When discussing our results in the next sessions, we always point out where and when supernovae are expected. The extent of which the supernova ejecta stays trapped to mix with the gas, remains a question. It may as well depend on the nature of the explosions themselves, if they are weak (`failed') or strong (`successful') supernovae, a question currently undergoing some investigation \citep{MacFadyen:1999,OConnor:2011,Smartt:2015}. 
Also, it remains a question how the supernova feedback influences star formation of the second generation stars. It may enhance it or stop it; with our current method, we have no ways to know. So, all these questions around supernovae are left to be investigated in future work.

	
	\section{Cluster wind and secondary star formation}\label{sec:Clusterwind}

	\begin{figure*}[t!]
		\centering
		\includegraphics[height=.99\columnwidth,angle=270]{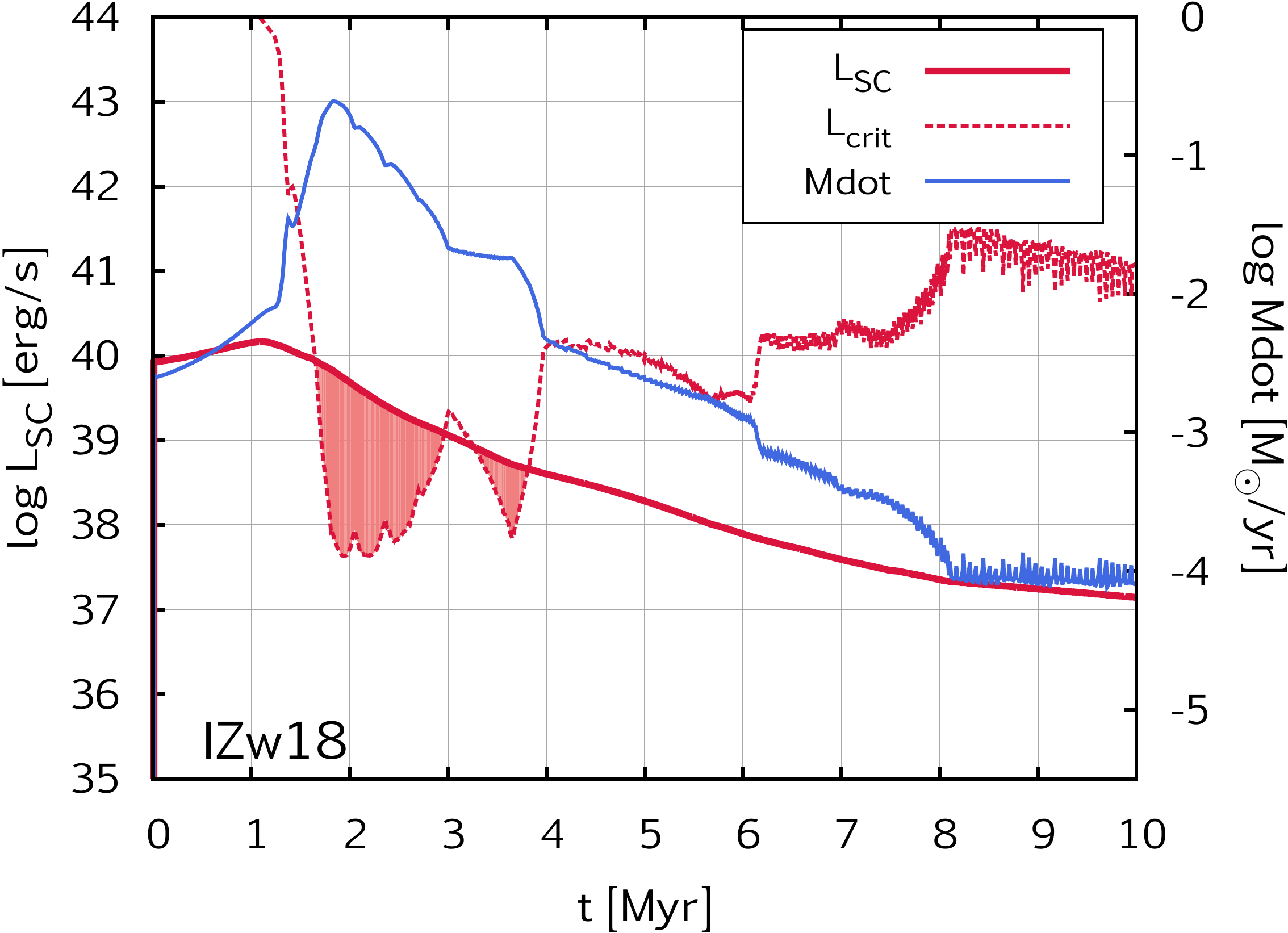} \hspace{10pt}
		\includegraphics[height=.99\columnwidth,angle=270]{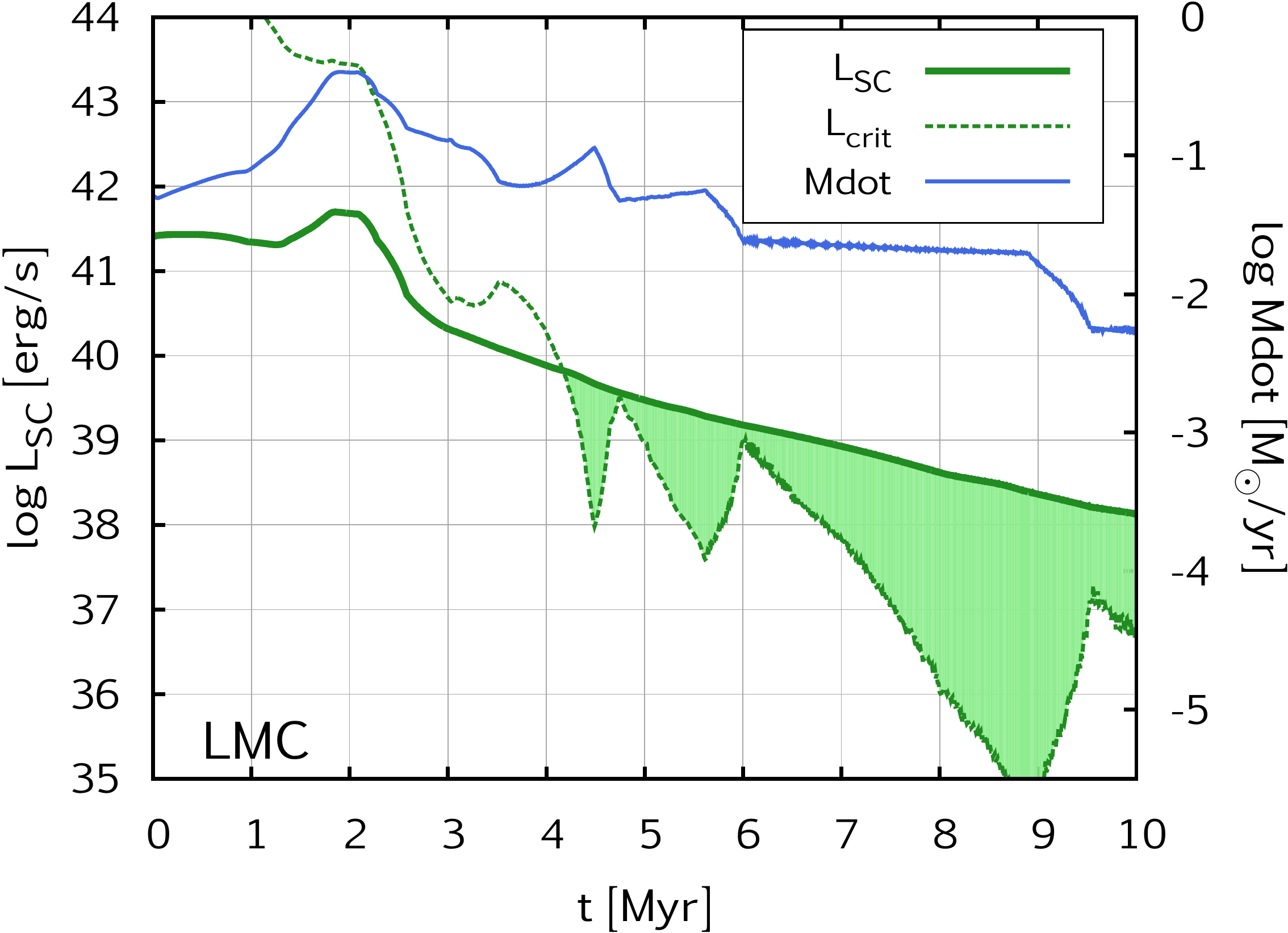} 
		\caption{Time evolution of the cluster wind. Whenever the cluster wind luminosity L$_{\mathrm{SC}}$ exceeds the critical luminosity L$_{\mathrm{crit}}$, wind mass is supposed to accumulate into the cluster center (marked by the shaded regions). Mass loss rate of the stellar populations is also shown (blue lines, values on the rigth axis). In the I~Zw~18 population, mass accumulation happens early on: before the first supernova explosions happen at 4~Myr. Thus we conclude that this early mass accumulation episode provides a `window' for the undisturbed formation of a second generation of stars. 
		}\label{fig:lum}
	\end{figure*}	

	\begin{figure*}[t!]
		\centering
		\includegraphics[height=.99\columnwidth,angle=270]{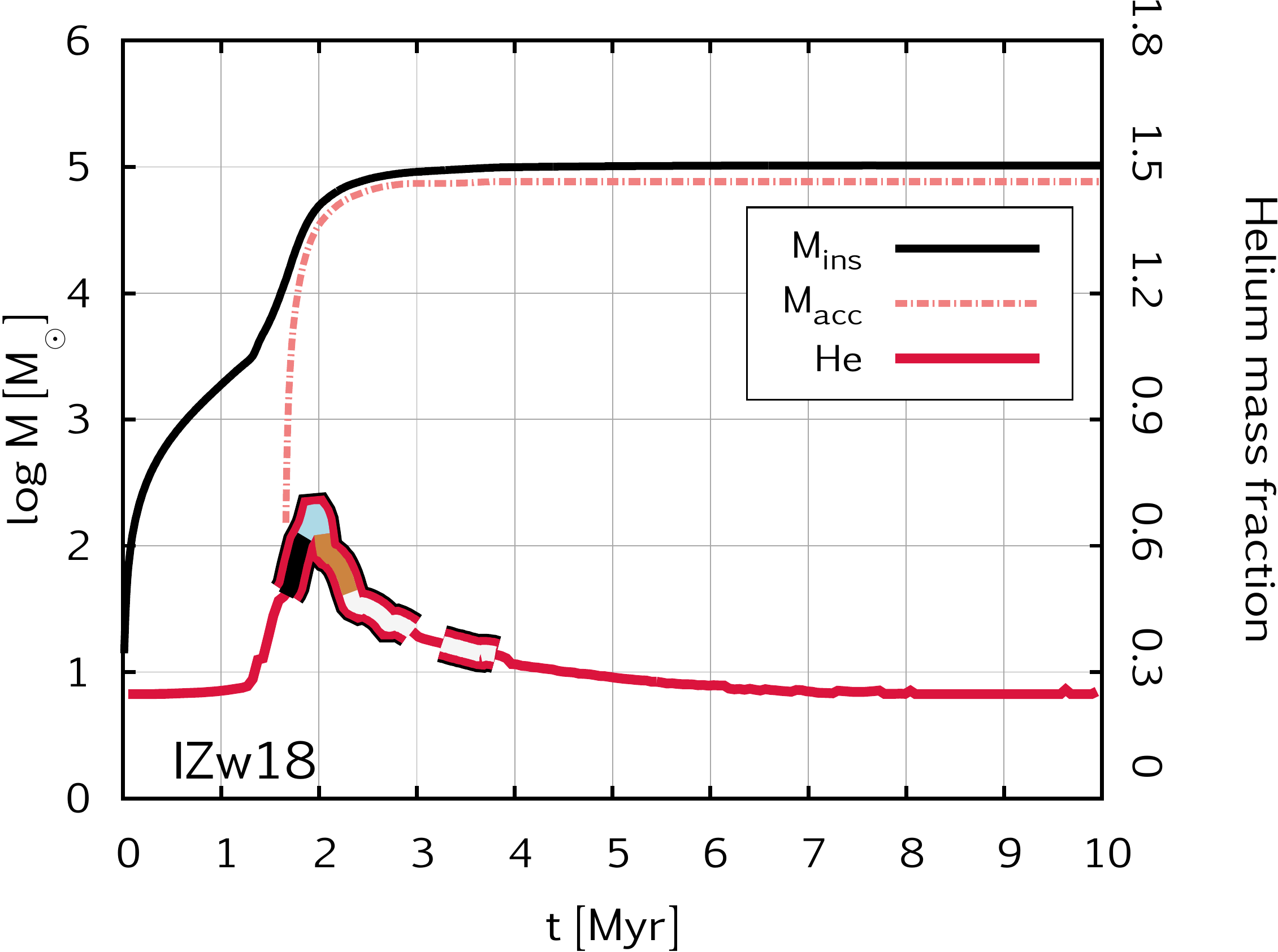} \hspace{10pt}
		\includegraphics[height=.99\columnwidth,angle=270]{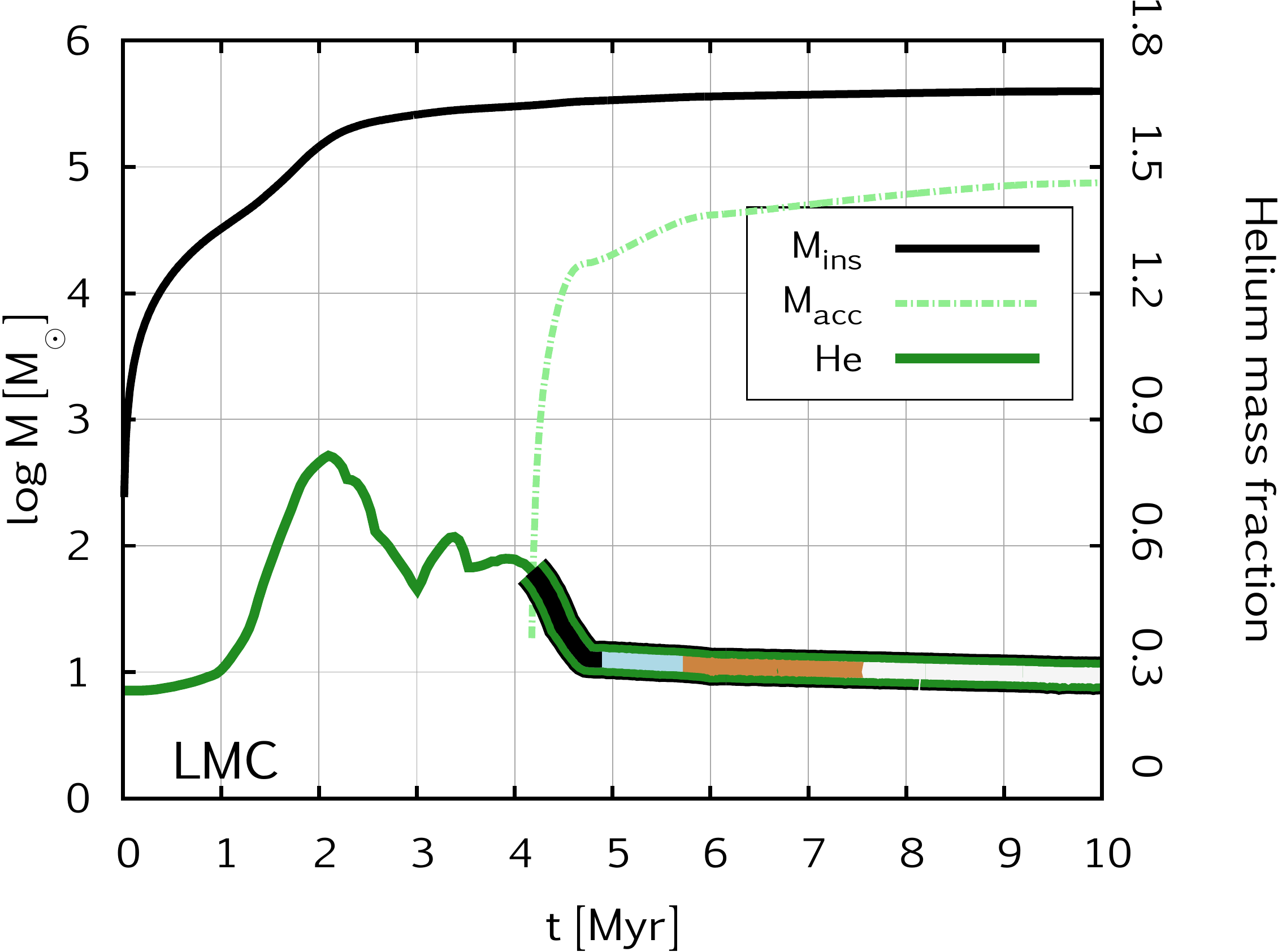} \\
		\caption{Time evolution of the mass lost from massive stars (i.e. inserted into the cluster wind by stellar winds, M$_{ins}$ and the mass which is accumulated into the cluster center (M$_{acc}$), both on a cumulative scale (left axis). Also shown is the helium mass fraction (right axis).  The periods when mass is actually accumulated (cf. the shaded regions in Fig.~\ref{fig:lum}) are indicated with thick lines, the four colors (black, blue, brown, white) corresponding to every 25\% of the total accumulated mass. 
		}\label{fig:acc}
	\end{figure*}

	As explained in Sect.~\ref{sec:windcalc}, mass accumulation happens when the cluster luminosity L$_{\mathrm{SC}}$ is lower than a certain critical luminosity L$_{\mathrm{crit}}$. The former is simply the combined mechanical energy of all stellar winds, while the latter is determined by the hydrostatic structure of the cluster (computed in our semi-analytical model) and takes into account various effect such as cooling and the stability of the cluster wind. Mass accumulation means that mass is removed from the wind and supposed to accrete onto forming proto-stars of a secondary generation---given of course that its velocity, inherited by the gas' velocity, is lower than the escape velocity from the cluster's gravitational potential.
	
	Figure~\ref{fig:lum} shows how the cluster luminosity L$_{\mathrm{SC}}$ relates to the critical luminosity L$_{\mathrm{crit}}$ over the calculations. The phases when L$_{\mathrm{SC}}$ exceeds L$_{\mathrm{crit}}$ are are shaded in the figure; however, the size of the shaded area should not be taken as indicative of anything, especially not the efficiency of star formation; it only serves to help us see when and how long the star formation is going on. 
	
	\subsection{Mass accumulation at low-metallicity}\label{sec:Macc-lowZ}
	
In the case of our low-metallicity stellar population, mass accumulation starts at $\sim$1.6~Myr and lasts until $\sim$3.9~Myr, resulting in a star formation episode that is $\sim$2.3~Myr long. 	
Figure~\ref{fig:lum} also shows the total mass loss rate of the cluster---that is, the mass loss rate of all stars in the population combined. The mass accumulation episode coincides with a pronounced peak in the mass loss rate; but this is not really a coincidence, as both effects are caused by the presence of cool supergiants in the population. Indeed, as discussed in Sect.~\ref{sec:grids}, massive and very massive supergiants are present in the low-metallicity population (but are absent in the high-metallicity population). These stars have low surface temperature (thus a slow stellar wind) and a high mass loss rate. Therefore, they facilitate star formation by \textit{not} heating the gas too much since they eject material with a small velocity and thus keep the cluster wind velocity also rather low. Additionally they deposit a huge amount of mass into the cluster in just the right time for it to accumulate in the center. 
	
	As discussed in Sect.~\ref{sec:supernovae}, the first core-collapse supernova explodes when the low-metallicity cluster is about 4.5~Myr old. Mass accumulation in this cluster happens before that age. Thus we conclude that in our low-metallicity calculations, \textit{the proto-stars of the second generation will have been already formed out of the stellar wind material before iron is deposited into the gas via supernova explosions.} 
	
	\textbf{Another caveat that this cluster avoids due to accumulating mass early, is that observationally, YMCs typically remove their gas by $\sim$4~Myr \citep{Hollyhead:2015}}. 
	Figure~\ref{fig:acc} shows the mass that is lost in stellar winds (i.e. inserted in the cluster by stars, M$_{ins}$) as well as that accumulated in the center (M$_{acc}$). The mass starts to accumulate when L$_{\mathrm{crit}}$ first exceeds L$_{\mathrm{SC}}$ (i.e. at $\sim$1.6~Myr). Almost all the mass that is inserted into the cluster wind accumulates in the center. \textbf{When the process ends (at $\sim$3.9~Myr)}, the total accumulated mass is almost 10$^5$~M$_{\odot}$. This is the mass budget from which a second generation of stars form. 
	
	
	\subsection{Mass accumulation at high-metallicity}\label{sec:Macc-highZ}
	
	In the case of our high-metallicity stellar population in Fig.~\ref{fig:lum}, mass accumulation starts at a later time than at low-metallicity, at 4.2~Myr. After this time, L$_{\mathrm{SC}}$ exceeds L$_{\mathrm{crit}}$ and keeps exceeding it until the end of our calculation. 
	
	In this case however, we do not imply that a second generation of stars should be expected to form out of the accumulated mass. As mentioned in Sect.~\ref{sec:supernovae}, this population experiences supernova explosions starting at the age of 2.1~Myr. Thus, our calculation should not be taken on face value after 2.1~Myr. We can nonetheless draw some interesting conclusions from it.
	
	The reason this population does not produce a mass accumulation episode as early as the low-metallicity population, is that the very luminous supergiants are practically absent. In this population, stars above 40~M$_{\odot}$ become LBVs or WR~stars (as discussed in Sect.~\ref{sec:grids}) which are hot stars with fast winds. So, while the mass deposited into the gas from the stellar winds is high due to the high mass loss rates of LBVs and WR~stars (higher than in the low-metallicity population at any given point in time), the cluster wind luminosity does not exceed the critical value until the first red supergiants, those with $\sim$45-40~M$_{\odot}$ and below, appear. Indeed, the total lifetime of a 45~M$_{\odot}$ model is 4.2~Myr, marking the point where L$_{\mathrm{crit}}$~$>$~L$_{\mathrm{SC}}$ and the mass accumulation starts. 
	
	That is, if we disregard supernova explosions. It falls outside the scope of current work to investigate what happens to the supernova ejecta under the conditions in this cluster: whether it gets shocked and cools staying and mixing with the gas, or leaves the cluster; and whether it enhances or stops star formation. 
	What we can conclude from our calculation nonetheless, is that mass accumulation starts $\sim$2.5~Myr later at high-metallicity (i.e. in the absence of cool supergiants) than at low-metallicity (when their contribution dominates). In the latter case, we expect that the accumulated mass forms a second generation of stars; while in the former case, we cannot be certain if a second generation forms without conducting 3D hydrodynamic simulations of the supernova ejecta and its contribution to star formation in the cluster.

	
	\section{Chemical composition of the second generation}\label{sec:Composition}

	\begin{figure*}[t!]
		\centering
		\includegraphics[height=.99\columnwidth,angle=270]{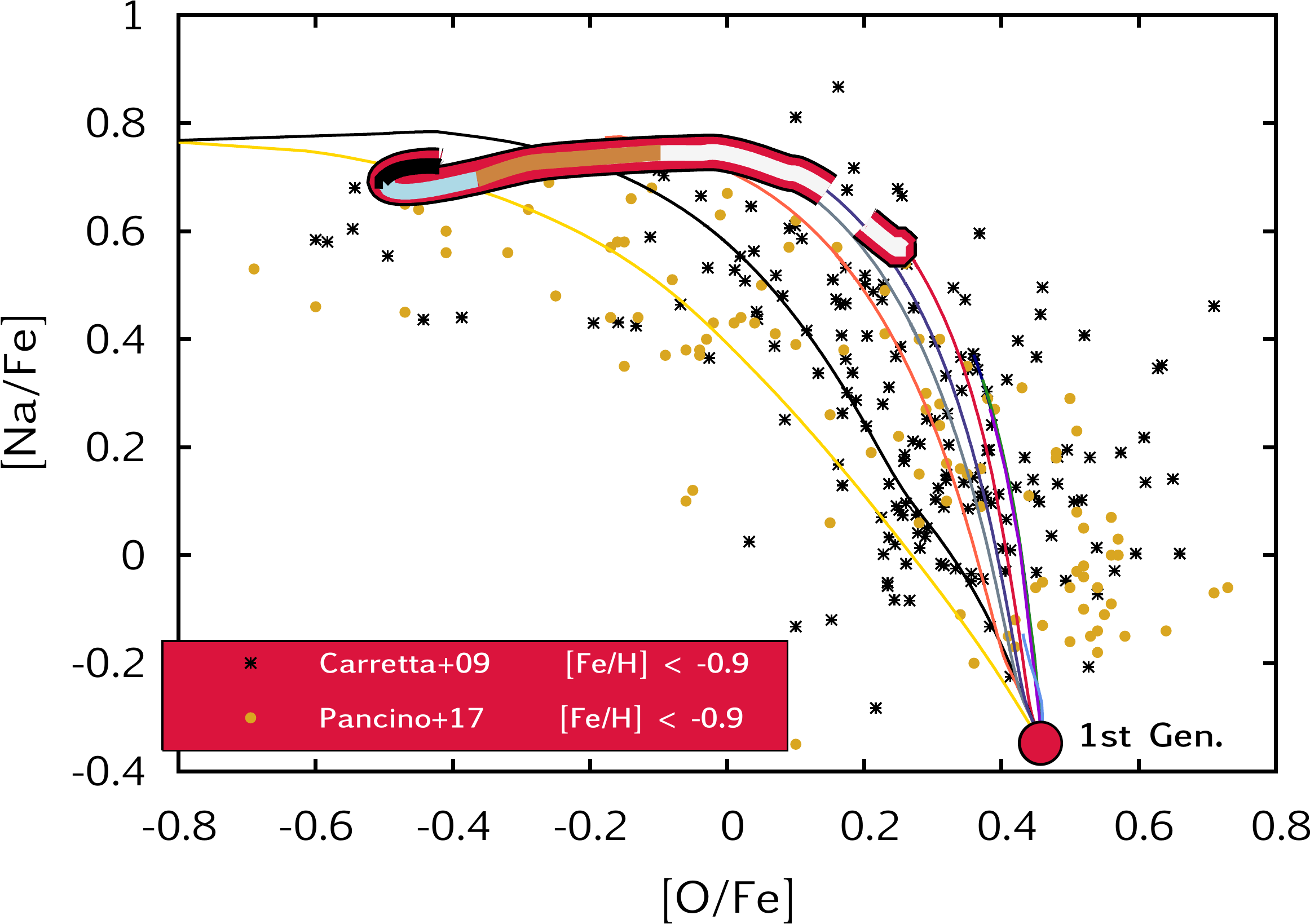} \hspace{10pt}
		\includegraphics[height=.99\columnwidth,angle=270]{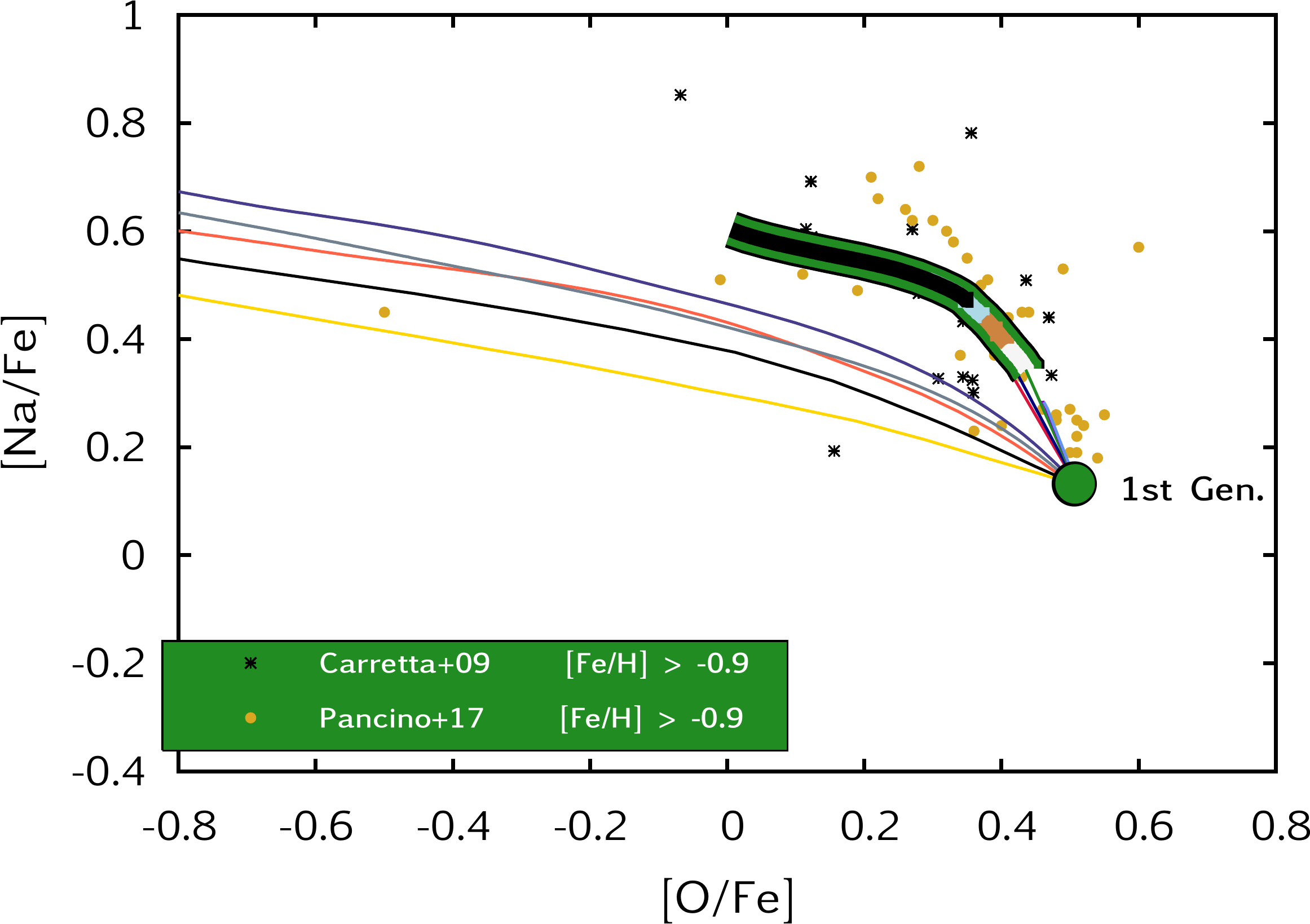} \\
		\vspace{10pt}
		\includegraphics[height=.99\columnwidth,angle=270]{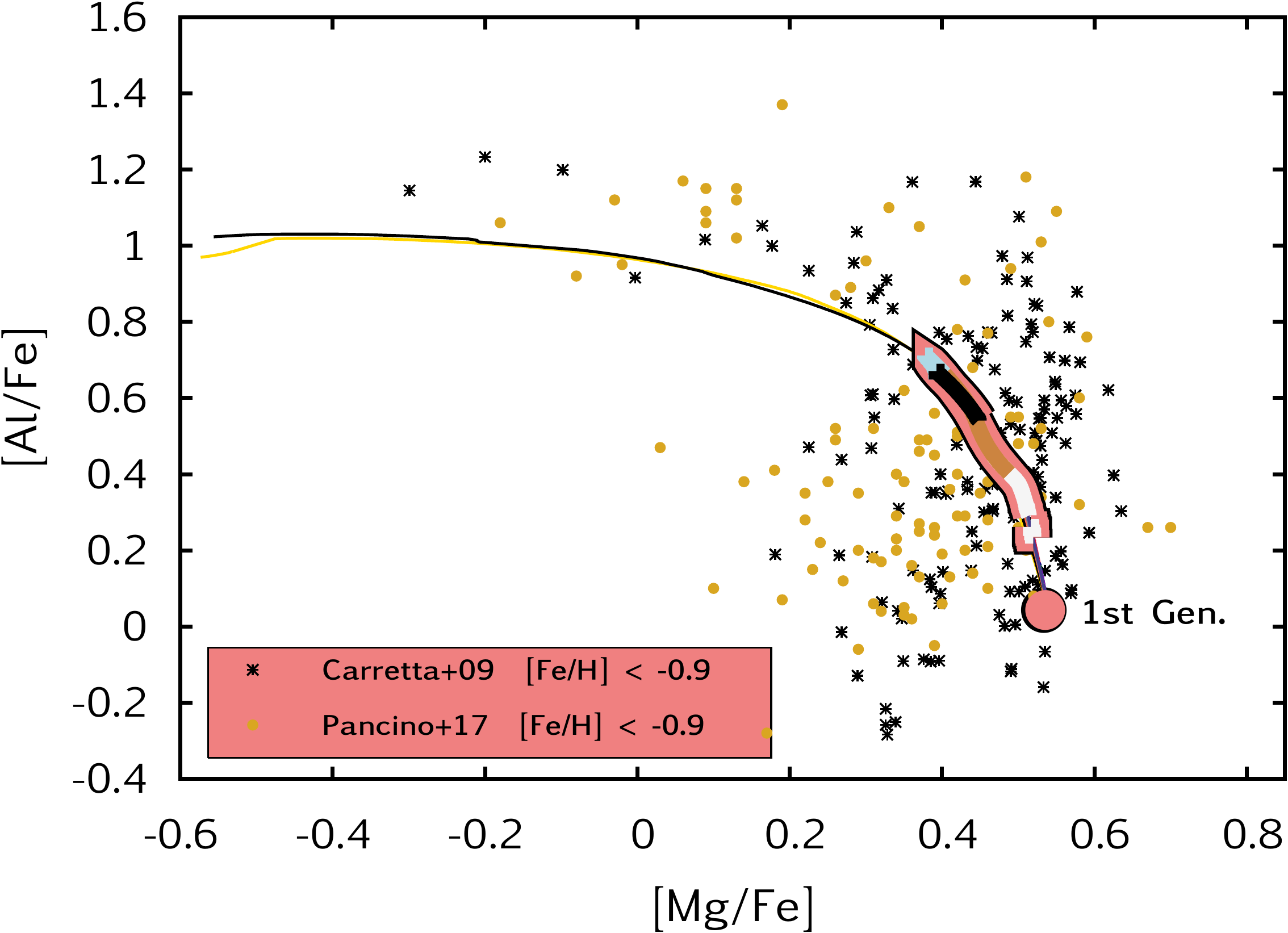} \hspace{10pt}
		\includegraphics[height=.99\columnwidth,angle=270]{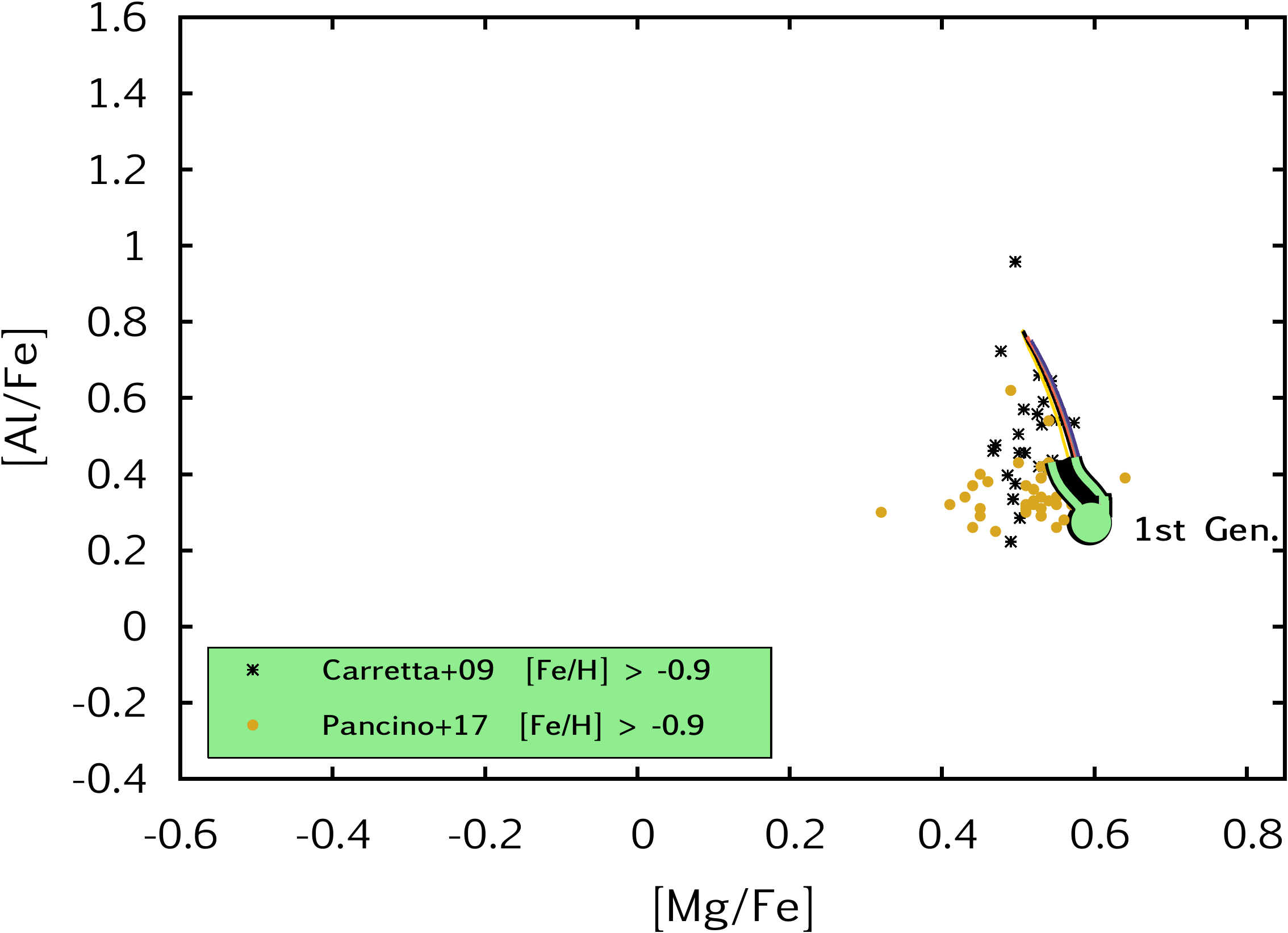} 
		\vspace{10pt}
		\caption{Surface composition of GC stars, showing the anticorrelations between oxygen-sodium and magnesium-aluminium. Observational data is taken from \citet[][FLAMES-UVES survey]{Carretta:2009a} \textbf{and \citet[][\textit{Gaia}/ESO-UVES survey]{Pancino:2017}.} The data corresponds to red giant stars in a total of 22 galactic GCs, with metallicities ranging from $-$0.9 to $-$2.3 on the left panels and from $-$0.4 to $-$0.9 on the right panels (in units of [Fe/H]). Observational error is typically between 0.05$-$0.12~dex. The composition of the pristine gas, that is, the composition we attribute to the first stellar generation in our calculation is marked by the filled circle (labelled `1st Gen.').
		Our calculation predicts that the mass accumulated in the cluster center (cf.~Fig.~\ref{fig:lum}) have a composition shown by the thick line. The four colored stripes overplotted on top of the thick line (black, blue, brown and white) mark the four quadrants of the total mass (i.e. every 25\%, starting with the black and ending with the white; cf. the same color coding in~Fig.~\ref{fig:acc}). Thin lines show the surface composition of the original stellar models during all their evolution; for the meaning of the colors, see Fig.~\ref{fig:HRDs}. We expect that the accumulated mass will mix with the original gas in the center. Thus, if a star forms out of it (as we suppose is the case for the low-metallicity cluster, cf. Sect.~\ref{sec:Macc-lowZ}) its composition will be a mixture of the pristine and the accumulated composition. 
		}\label{fig:abundance}
	\end{figure*}

	\subsection{Light element variations} 
	
	Light element abundance variations are a well-established observational fact for practically all GCs which have been extensively studied spectroscopically \citep[see e.g.][for a recent review]{Bastian:2018}. In particular, Na overabundance is always observed together with O depletion; while the sum of C, N and O is constant, suggesting that the CNO-cycle is operating. In some GCs, Al and Mg display variations as well. This implies that at least one population of low-mass stars (usually referred to as the second generation) is made up of material that has previously undergone hot-hydrogen burning. This process is known to be active only in massive and intermediate mass stars because low-mass stars' core temperatures are not high enough for that. 
    
    We do not study the contribution of intermediate mass stars (i.e. asymptotic giant branch, AGB, stars) to our cluster, since their contribution happens after $\sim$30~Myr (while our computation ends at 10~Myr). Nonetheless, AGB stars have been used to account for the observed light-element variations, see e.g. \citet{Cottrell:1981,Karakas:2006,DErcole:2010,Doherty:2014}. Here we focus only on massive stars, but point out that AGB~stars may contribute to the cluster gas' composition at later ages. 
    
	Figure~\ref{fig:abundance} shows the observed anticorrelation between two pairs of light elements, O vs. Na and Mg vs. Al. The data is taken from \citet{Carretta:2009a} which is a FLAMES-UVES survey of $\sim$200 red giant stars in 17~GCs; \textbf{as well as from \citet[][and private comm.]{Pancino:2017} which contains $\sim$150 red giant stars in 9~GCs with \textit{Gaia}/ESO-UVES abundance measurements of all four elements in question. Since the abundance scales used by these two surveys differ, we took into account a small (compared to the internal spreads) offset of $\sim$0.10$-$0.15~dex when plotting the two data sets next to each other, as suggested by \citet[][and private comm.]{Pancino:2017}.}
	
	We account for low-metallicity and high-metallicity clusters in a simple way, by dividing the observational samples into two categories: \textbf{one with [Fe/H]~$>$~$-$0.9 and another with $<$~$-$0.9. The choice of this value is motivated by Fig.~3 of \citet{Harris:2010} in which the metallicity histogram of a large catalogue of GCs seems to show two peaks, with an arbitrary division at around~$-$0.9. In future work this has to be refined by investigating a range of different metallicities, as discussed in Sect.\ref{sec:future}.}
	According to Fig.~\ref{fig:abundance}, both anticorrelations are observed to be much more pronounced at low-metallicity than at high; with no significant Mg depletion found amongst high-metallicity clusters whatsoever in these data sets. 
	
	When comparing our theoretical predictions to the observed spreads, we suppose that our low-metallicity cluster (with [Fe/H]~$=$~$-1.7$) is representative for all GCs below $-0.9$ and the same for our high-metallicity cluster (with $-$0.4) for above $-$0.9.
	As the stellar models do not use an $\alpha$-enhanced mixture \citep[as suggested for GC stars by e.g.][see their Table~3]{Decressin:2007} but a mixture suitable for certain galaxies, therefore when creating Fig.~\ref{fig:abundance}, the initial O, Na, Mg, and Al abundances of our models are scaled to match the composition of the unpolluted red giants.
	Below we discuss what our calculations predict for the composition of the second generation of stars (or more precisely, as explained in Sect.~\ref{sec:Macc-highZ}, for that of the mass accumulated in the cluster center, out of which a second generation forms at low-metallicity). 
	
	\subsubsection{Na \& O at low-metallicity}\label{sec:NaO}
	
	Our calculation of a low-metallicity cluster predicts that the mass accumulated in the center has a high sodium value and a large range of oxygen values. In fact, about half of the accumulated mass has extremely low oxygen abundance with high sodium (cf. black and blue stripes), while another half is spread out in oxygen. This is related to \textit{when} the mass is accumulated: if it is accumulated early, its composition is dominated by the winds of the most massive supergiants (which evolve to the supergiant branch earlier); while if late, it is dominated by the less massive ones. 
	
	It is expected that in the center, this material mixes with the pristine gas (out of which the first generation of stars formed). Thus, the mixture of the accumulated mass and the original gas can possibly produce the whole observed range of Na-O abundances in stars of low-metallicity clusters.
	
	\textbf{Such dilution of polluted gas with pristine gas is typically also invoked by other scenarios \citep[e.g. in Eq.~7 of][]{Decressin:2007b}. One caveat here is that observationally, YMCs have removed their gas by $\sim$4~Myr \citep{Hollyhead:2015}, which presents a challenge for all models that form the second generation after this age. But in our low-metallicity cluster governed by the presence of supergiants, this caveat is avoided by the `window' for starformation opening between 1.6$-$3.9~Myr (as shown in Fig.~\ref{fig:lum}).}
	
	\subsubsection{Na \& O at high-metallicity}
    
    Metal-rich clusters show a smaller extent of both Na and O variations than metal-poor ones, with the lowest observed Na value being about 0.4~dex higher. As explained by \citet{Carretta:2009a} this is because the plateau of minimum Na established by (a previous generation) of supernova nucleosynthesis is a function of the metallicity. 
    
    Although we cannot infer from our calculation of a high-metallicity cluster that a second generation of stars would form from the accumulated mass (due to the uncertainties associated with supernova explosions, cf. Sect.~\ref{sec:supernovae}), it is nonetheless interesting to compare our predictions to observations of high-metallicity cluster stars in Fig.~\ref{fig:abundance}. 
    As opposed to the low-metallicity case discussed above, now the most massive stars' mass loss have no contribution to the accumulated mass, as mass accumulation starts when these are already dead. The composition in this case is dominated by supergiants of initial mass below 40~M$_{\odot}$. They do produce some Na by destroying some O, but much less of an extent than higher mass stars do, especially when it comes to destroying oxygen. This is in accordance with the observational data in Fig.~\ref{fig:abundance}.   
    
	\subsubsection{Al \& Mg at low-metallicity}\label{sec:MgAl-lowZ}
    
    This is the case where our calculation struggles to account for the whole extent of observations. While red giants in low-metallicity GCs display a broad spread in both Mg and Al, our prediction is that the second generation of stars would have an Al spread only about 0.6~dex with almost no Mg being destroyed. This is so even though our most massive models do lose material with a very low Mg abundance (cf. yellow and black thin lines corresponding to stellar models with 257 and 575~M$_{\odot}$). 
    
    The reason for this lies in the specifics of the star formation episode. We accumulate mass in our calculation when the hydrodynamical conditions in the cluster are just right (cf. Sect.~\ref{sec:Macc-lowZ} and Fig.~\ref{fig:lum}). This means that we have a star formation episode that lasts from 1.6~Myr to 3.9~Myr, during which the material lost by massive stars is integrated together (with properly weighting by the IMF) to produce the stripes in Fig.~\ref{fig:abundance}. Our very massive models gradually lose their outer layers via their stellar winds, starting with those layers that are less effected by hot-hydrogen burning. The first stars of the second generation form out of this material (black stripe). When these very massive stars are already losing their deeper, magnesium-deficient layers, lower-mass stars have evolved to the supergiant branch, contributing to the total composition significantly. The material of their surface layers is, therefore, accumulated together with the deeper layers of the very massive stars, resulting in some slight decrease in the combined abundance (blue stripe) but clearly not enough to account for the whole observed spread in Mg. The later phases of star formation, when even lower mass supergiants dominate the composition, produces a second generation with an even higher Mg (yellow and white stripes).
    
    Nonetheless, all the extremely low Mg values were observed in the same cluster, NGC~2808 \textbf{(in both data sets)}. 
    As discussed by \citet{Carretta:2009a}, there is a quite significant cluster-to-cluster variation when it comes to Mg and Al (see their Fig.~6), with some clusters displaying large, and some displaying small, spread of these elements. \textbf{Indeed, the phenomenon of magnesium depletion is not a common feature among GCs, rather an exception, with extremely low magnesium abundances only observed in a handful of clusters (e.g. NGC~2419, NGC~2808) and even there the situation is further complicated by the phenomenon's apparent dependence on mass and metallicity \citep[cf.][]{Pancino:2017}. }
    
    Our calculation is way to simplistic to account for this cluster-to-cluster variation, as we use only two sets of single stellar models at two given metallicity values, together with some---reasonable, but certainly improvable---assumptions about the secondary star formation. We discuss ways to improve our theory in Sect.~\ref{sec:future}. 
    
    It is, for example, quite conceivable that some stars do form out of the pure material (that is, mixed neither with the other stars' ejecta nor with the pristine gas) of the most massive supergiants. Such a scenario was suggested by \citet{Szecsi:2018} to possibly operate in the case of some very massive supergiants. 

	\subsubsection{Al \& Mg at high-metallicity}
		
	The high-metallicity models do not show the Mg-Al anticorrelation. Not even in the most massive case (up to 500~M$_{\odot}$). The reason for this is related to their core temperatures. To destroy magnesium and produce aluminium, the $^{24}$Mg(p,g)$^{25}$Al~chain should be active, which happens at a core temperature of $\sim$80$-$100~MK, according to \citet{Ventura:2011}. Below and above this temperature range, the reaction rate of the $^{24}$Mg(p,g)$^{25}$Al~chain is too small for it to produce any effect in stellar models. 
	
	But having the correct core temperature is not enough. In order to destroy a lots of Mg, the right thermodynamical conditions should last for a long time \citep[because the chain first creates $^{24}$Mg~$\rightarrow$~$^{25}$Mg, and then \textit{slowly} destroys
	$^{25}$Mg too; see Fig.~1 of][]{Ventura:2011}. So we need stars that not only \textit{pass through} the correct temperature range while, for example, collapsing or re-structuring, but keep burning their fuel with exactly the right temperature \textit{for a long time}. The longer the time the core has the right temperature, the more Mg is destroyed and converted into Al. 
	
	This differentiates between the LMC models and the I~Zw~18 models. For example, our LMC~model with M$_{ini}=$260~M$_{\odot}$ has T$_c<$~55~MK during almost all its main sequence lifetime. It only reaches the range 80$-$100~MK when the core contracts to ignite helium, but this is a rather short phase in the star's life ($\lesssim$10~kyr), after which the temperature increases way above 100~MK. The I~Zw~18 model with 257~M$_{\odot}$, on the other hand, already starts its evolution with 60~MK and then slowly increases to 75~MK. Although the literature cites 80~MK as the nominal lower limit for the reaction to be effective, we find in our models that already at $>$65~MK there is significant Mg depletion and Al production if this temperature lasts for a long time (in our 257~M$_{\odot}$ model, for $\sim$0.6~Myr).
    This is in accordance with observations (Fig.~\ref{fig:abundance}), which show that high-metallicity clusters have no significant variation in Mg. 
	
	\subsection{C, N, O and He abundances}\label{sec:CNOHe}
    
Observations of GCs typically show that carbon, nitrogen and oxygen abundances are in accordance with the CNO-cycle's equilibrium values. It means that the sum of these three atoms is constant; they simply act as catalysts in the cycle. Nonetheless, C and O drops and N increases in later populations due to the CNO-equlibrium values being different from the abundances of the original gas. 

This is confirmed by our theoretical calculations. Our massive stellar populations (both at low- and at high-metalicity) do conserve the sum of C, N and O, with their respective abundances being consistent with the CNO-equilibrium values. 
\textbf{
However, as the available C and N data for the first generation stars are sparse and the interpretation of C and N variations is complicated by some evolutionary effects in the red giant branch phase \citep[e.g.][]{Boothroyd:1999,Gratton:2000}, we refrain from fitting the C~\&~N anticorrelation here. }
    
The helium mass fraction of the accumulated mass in our calculation is shown in Fig.~\ref{fig:acc}. It reaches a much higher value than what is inferred from observations of any GC  \citep{Bastian:2015,Bastian:2018,Milone:2017}. We discuss the implications of this finding in Sect.~\ref{sec:heliumproblem}. Colored stripes in Fig.~\ref{fig:acc} represent the same as in Fig.~\ref{fig:abundance}, that is, the four quadrants of the mass accumulated in the cluster center. Comparing these two figures, we find that the most extreme oxygen depletion is produced together with a helium mass fraction, Y, ranging between 0.5$-$0.7 (i.e. black and blue stripes). This composition is a mixture of the material ejected from our most massive stars down to 150~M$_{\odot}$. 
However, a less extreme oxygen depletion is possible to reach with a helium mass fraction between 0.3$-$0.5 (white stripe). This happens at end of the star formation episode when the last 25\% of the mass is accumulated. This mass is made of the winds of supergiants with an initial mass 40$-$60~M$_{\odot}$. 
For further discussion on this issue, we refer to Sect.~\ref{sec:heliumproblem}.

No helium-burning products, nor products of later burning phases, are found in our calculations at any metallicity.

	
	\section{Mass budget}\label{sec:Massbudget}

\begin{figure*}[t!]
\centering
\includegraphics[width=1.4\columnwidth]{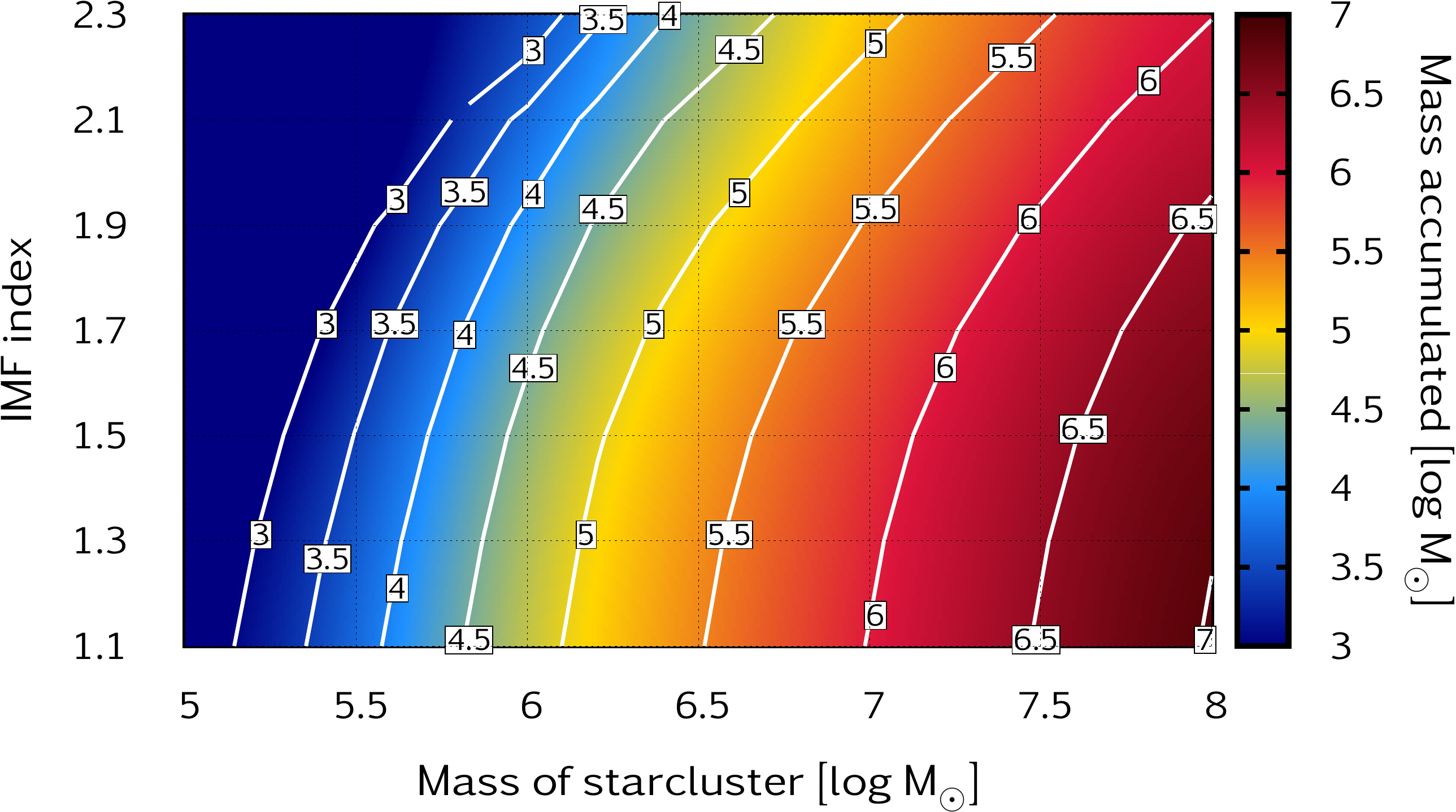}
\caption{Results of our parameter space study. We vary the initial mass of the cluster, M$_\mathrm{SC}$, shown on the X~axis, and the index of our IMF shown on the Y~axis. Contours present the outcome of our calculation in terms of the amount of mass that is accumulated in the cluster center. \textbf{Since the amount of accumulated mass (as well as the number of second generation stars, cf.~Fig.~\ref{fig:PS-N12}) is an increasing function of the initial mass, we suggest that this may explain why we observe a correlation between today's GC masses and the extent of a polluted second generation.}
}
\label{fig:PS-Macc}
\end{figure*}

\subsection{On the mass budget problem and dynamical removal of stars}\label{sec:dynamical}

The fraction of stars with anomalous chemical composition, i.e. the second generation, varies in the range of $\sim$30$-$90\% among GCs, with a mean value around $67$\,\% \citep{Milone:2017, Bastian:2018}. A difficulty of most models to predict such a high fraction of second generation stars is called the mass budget problem.

In our calculation of a low-metallicity cluster with a total initial mass of 10$^7$~M$_{\odot}$, we find that about 10$^5$~M$_{\odot}$ is available to form second generation stars. We emphasise that this is the mass that is ejected from massive stars via their winds and accumulated inside the cluster center---both processes accounted for in our calculations---and not, for example, the total mass in these massive stars. We took into account a standard IMF with an upper mass limit of 500~M$_{\odot}$. As opposed to earlier works accounting for the mass accumulation process \citep[e.g.][]{Wunsch:2017}, we did not include additional ad-hoc parameters such as mass-loading or heating efficiency, but found that the presence of supergiants already facilitates the process.

Globular clusters today have a typical total mass of a few times 10$^5$~M$_{\odot}$. If they indeed used to be massive clusters born with 10$^7$~M$_{\odot}$, they must have lost $\gtrsim$90\% of their mass during their lives. It has been suggested \citep{DErcole:2008,Decressin:2010,Vesperini:2010,Khalaj:2015} that clusters may lose a huge fraction of their stars via dynamical `evaporation'. This process mainly effects stars located in the outskirts of the cluster.

In our calculations, mass accumulation happens in the center of the cluster\footnote{As shown by \citet{Wunsch:2017}, the outcome of the semi-analytic code we use here is in accordance with 3D hydrodynamic simulations, which show that the cool gas falls towards the cluster center to accumulate there.}. This means that even if mass is removed from the cluster very efficiently by dynamical evaporation, the $\sim$10$^5$~M$_{\odot}$ proto-stars of the second generation would be very hard to eject via this process, due to them being centrally located. 
For example, N-body simulations of the cluster's long-term dynamical evolution carried out by \citet{Khalaj:2015} show that it is indeed possible to explain present day observations by a cluster that contained a second generation number fraction of 10\% initially, on the conditions that a substantial amount of gas is kept after the formation of the second generation, and that this gas is then expelled on a very short timescale. 
\textbf{If such a process takes place in the cluster leading to the loss of almost only first generation stars, our low-metallicity model presented in Fig.~\ref{fig:acc} is able to fulfill the mass budget, having already accumulated and converted $\sim$10$^5$~M$_{\odot}$ into second generation stars.}

\textbf{However, it is not conclusively established that the dynamical evolution leads to the loss of only first generation stars. A recent study \citep{ReinaCampos:2018} suggests that while GCs have indeed lost 90$-$95\% of their initial masses, the present-day ratios of first vs. second generation reflect the initial values. 
Another argument comes from \citet{Kruijssen:2015} who suggests that if mostly first generation stars were lost due to tidal interactions with the host galaxy, we should expect to observe GCs with increasing mass-loss towards smaller galactocentric radii, with higher gas pressures at birth and with higher cluster metallicities (cf. Sect.~2.1.2 and especially argument~(v) on page~1661 of the cited paper). 
While it could be insisted that the measurements of these quantities are significantly impacted by uncertainties, it is nonetheless clear that there are far too many open questions regarding the dynamical evolution during which YMCs become GCs for it to be called an established theory.}

\textbf{We have no means of solving any of these open questions here, due to us only focusing on the relatively short term phase of starformation.}
Indeed, our calculation only involves the first 10~Myr of the cluster's life, as we are mainly interested in the mass accumulation process. \textbf{We predict that the mass is accumulated in the cluster center and the second generation of stars form there. Still, we cannot directly quantify which fraction of the first generation would be lost over the subsequent lifetime; nor, therefore, the final ratio of first vs. second generation we may expect in today's GCs after them having been undergone several gigayears of dynamical evolution.}

We can nonetheless \textbf{provide upper limits and predict some trends,} by studying how our calculation is affected if we vary initial conditions such as the total mass of the cluster at birth or the IMF. This is done in what follows. \textbf{We emphasize that from now on, we do not suppose that the dynamical ejection process prefers the first generation. If it does, it helps our case; but there are other options too to alleviate the mass budget problem.}

	\subsection{Parameter space study}\label{sec:paramspace}
  
We repeated our low-metallicity calculation with varying two of the input parameters, the total initial mass of the cluster, M$_\mathrm{SC}$, and the high-mass index of the IMF, $\alpha_4$ (our IMF is explained in Sect.~\ref{sec:popsyn}).
We vary M$_\mathrm{SC}$ between 10$^5$..10$^8$~M$_{\odot}$ and $\alpha_4$ between $-$1.1..$-$2.3.
The results are summarized in Fig.~\ref{fig:PS-Macc} (and Sect.~\ref{sec:PS-Macc}) for the amount of accumulated mass and in Fig.~\ref{fig:PS-N12} (and Sect.~\ref{sec:PS-N12}) for the number ratio of the second generation stars vs. the total, $N_2/(N_1+N_2)$. 
 Our motivation for testing the high-mass index of the IMF is that recently, some attention has been paid to measuring this value in various star forming regions with various methods. Some of these works report the finding of a top-heavy IMF \citep{Kalari:2018,Schneider:2018,Zhang:2018}, although the debate seems not to be over \citep{Bastian:2010,Khorrami:2016, DeMasi:2018,Farr:2018,Hopkins:2018,Schneider:2018b}. 
 
 \subsection{Mass accumulation as a function of initial cluster mass and IMF}\label{sec:PS-Macc}
  
We find that the amount of accumulated mass is an increasing function of both the total initial mass and the number of massive stars in the population. Clusters in the top left corner of Fig.~\ref{fig:PS-Macc} do not accumulate a noteworthy amount of mass; in the case of a cluster with M$_\mathrm{SC}$~$=$~10$^5$~M$_{\odot}$ and a classical IMF index, no mass accumulation is taking place. This suggests that at least the lowest mass GCs with multiple populations must have started out as more massive clusters and then gotten rid of some of their mass over their lives. 

In the most extreme, probably hypothetical case of a 10$^8$~M$_{\odot}$ cluster with an extemely top-heavy IMF, as much as 10$^7$~M$_{\odot}$ is available to form the second generation. \textbf{The consequences of this are discussed further in Sect.~\ref{sec:morethantwo}.}

As for the chemical composition, we find the following. The \textbf{\textit{total} spreads predicted} in Na and O, as well as those in Mg and Al, together with the high helium abundance (Sect.~\ref{sec:CNOHe}), are practically unchanged when varying the initial parameters. \textbf{Nonetheless, more massive clusters produce more polluted gas, though with a less extreme \textit{average} composition. To be able to quantitatively compare these results to observations, we would need to model the mixing with primordial gas, which will be done in future work. Here we only establish a correlation between the initial cluster mass and the amount of (polluted) mass accumulating in the cluster center. What this correlation means in terms of number of stars, is discussed in the next sections.}


\subsection{The ratio of second vs. first generation stars}\label{sec:PS-N12}

\begin{figure*}[t!]
\centering
\includegraphics[width=1.4\columnwidth]{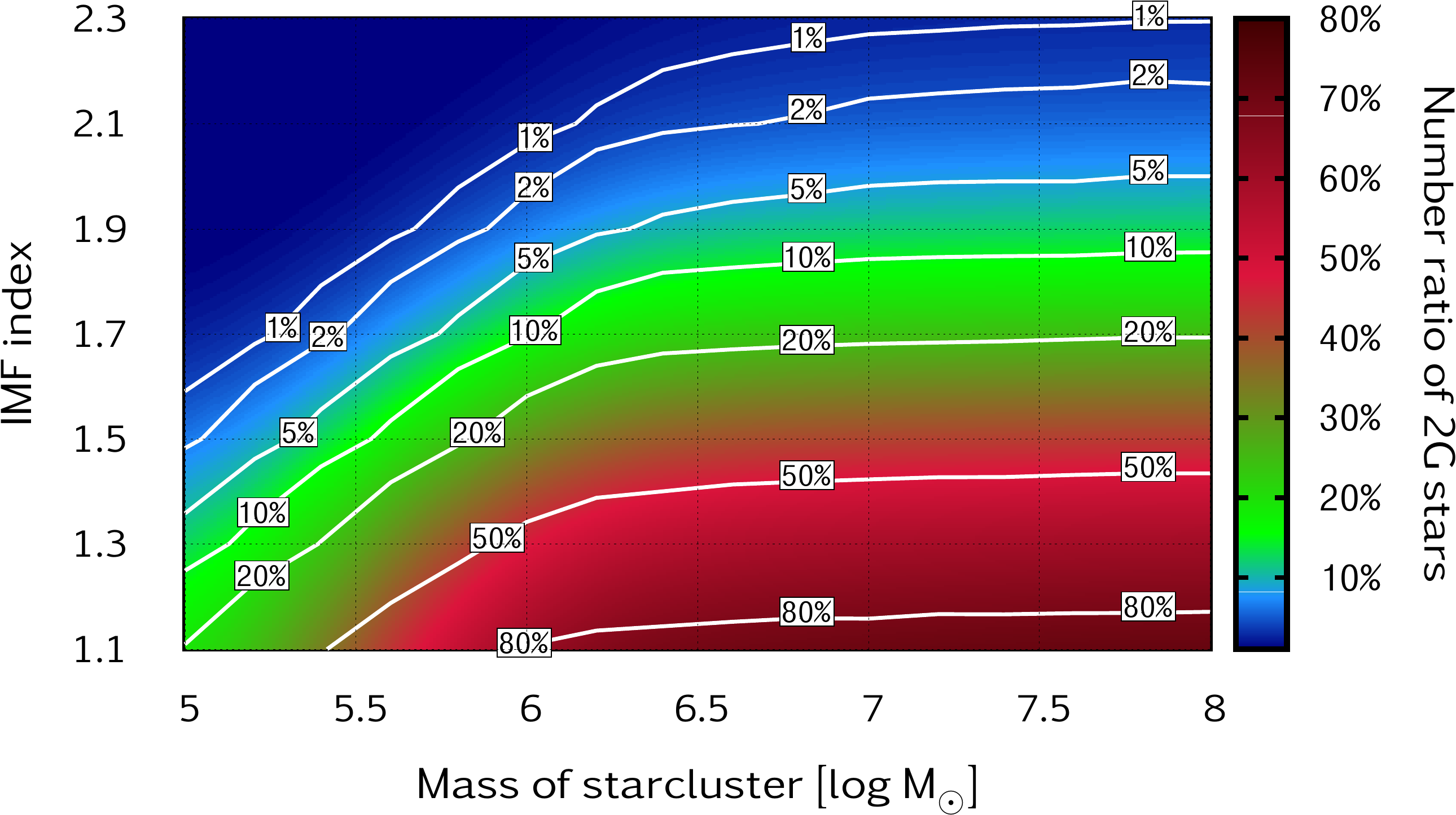}
\caption{The same as Fig.~\ref{fig:PS-Macc}, but for the number ratio of low-mass stars in the second generation (i.e. $N_2/(N_1+N_1)$, as explained in Sect.~\ref{sec:PS-N12}). Note that we do not suppose that only low-mass stars form in the second generation, but
apply a regular IMF with an index $-$2.3.}
\label{fig:PS-N12}
\end{figure*}

Figure~\ref{fig:PS-N12} shows the same parameter study as before, but here we include assumptions about the number of stars formed out of the available mass, as follows.
The number of low mass stars in the first generation, $N_1$, is taken to be stars with masses between $0.08$ and $0.8$\,M$_\odot$, as these are expected to be still alive after $\sim 12$\,Gyr of cluster evolution, and thus to be observed today. To calculate $N_1$ then, we assume a first generation with mass $M_\mathrm{SC}$ and a given IMF (depending on $\alpha_4$, with a mass range of $0.01 - 500$\,M$_\odot$). 
The number of low-mass stars in the second generation, $N_2$, is estimated from the amount of accumulated mass, $M_\mathrm{acc}$, by assuming that all this mass is used to form the second generation \textit{with a standard IMF}.
Similarly as before, we apply an IMF between $0.01 - 500$\,M$_\odot$, and take the number of stars in the mass range $0.08 - 0.8$\,M$_\odot$ to be our $N_2$. The ratio of generations plotted in Fig.~\ref{fig:PS-N12} is then $N_2/(N_1+N_2)$. 

We find that,
depending on the IMF index, the ratio of the generations can be anything between 1~and~80\%. 
We emphasize again that we do not suppose a second generation forming stars only up to 0.8~M$_{\odot}$ \citep[as done in e.g.][]{deMink:2009}, but apply a regular IMF which includes all stars, even massive and very massive stars. 
Additionally, the dynamical evolution of the star cluster should be taken into account: it may increase the $N_2/(N_1+N_2)$ ratio even further, as the second generation is expected to be more centrally concentrated (see the discussion in Sect.~\ref{sec:dynamical}). For observational evidence of a more centrally concentrated second generation, see e.g.~\citet{Milone:2012, Dalessandro:2016}.


Thus we conclude that either a top-heavy IMF or the dynamical removal of the non-centrally located stars---or a combination of these two effects---is our suggested explanation for the observations of the second generation being as populous as $\sim$30$-$90\%.

\subsection{On the correlation of second generation and cluster mass}

It has been reported by several authors that the fraction of enriched stars in GCs strongly correlates with the cluster mass. See e.g. Figs.~14 and~16 of \citet{Carretta:2010}, Fig.~20 of \citet[][]{Milone:2017} and Fig.~7 of \citet[][]{Carretta:2018}. 
To directly confirm these observations with our calculations, we would need to test the effects \textbf{of two processes: the mixing with pristine gas (cf. Sect.~\ref{sec:PS-Macc}) and} the dynamical removal of stars (cf. Sect.~\ref{sec:dynamical}). These tests fall outside the scope of the present study. However, we point out that both the total initial mass and the accumulated (i.e. enriched) mass do correlate with the number ratio of second generation stars in our calculations. Thus, unless the processes mentioned above cancel out this correlation for some reason (e.g. ejecting only centrally located stars; or removing more mass from the more massive clusters, thus making the initial cluster mass strongly anticorrelate with the present day cluster mass), \textit{we predict this trend to be indeed imprinted on GCs observed today}. 

\subsection{More than two generations?}\label{sec:morethantwo}

\textbf{In our model, we allow the second star formation episode to also create massive stars (see the discussion in Sect.~\ref{sec:PS-N12}). Thus, it is possible in principle that the whole scenario repeats another time. As seen in Fig.~\ref{fig:PS-Macc}, our most extreme cluster with M$_{\mathrm{SC}}$~$=$~10$^8$~M$_{\odot}$ can accumulate as much as 10$^6$~$-$~10$^7$~M$_{\odot}$ in the center---which may again form massive stars and thus repeat the cycle again. Without testing this possibility quantitatively here, we speculate that it may explain the fact that some clusters are observed to have more than two stellar generations \citep[e.g. NGC~2808, cf.][]{DAntona:2005,Milone:2015}.}

	
	\section{Discussion}\label{sec:discussion}
	
\subsection{On the helium problem}\label{sec:heliumproblem}

Direct helium abundance measurements are difficult to obtain spectroscopically, since the relevant photospheric transitions need a much higher surface temperature than what red giant stars (those we have extensive spectroscopic studies of when it comes to light elements; cf. the dataset plotted in Fig.~\ref{fig:abundance}) have. Helium abundance is usually inferred indirectly from photometric measurements via fitting stellar models (isochrones) in the observed color-magnitude diagrams. These studies suggest that almost all GCs show variation in helium abundance, albeit rarely up to Y$_{\mathrm{obs}}$~$=$~0.35.


In Sect.~\ref{sec:CNOHe} we show that our models overpredict helium at an extent (Y~$\sim$~0.5$-$0.7) that is not reconcilable with observations. However, as pointed out by several authors already \citep{Lochhaas:2017,Bastian:2018,Szecsi:2018}, this is a generic problem in the field. The current understanding of the nucleosynthetic origin of the second generation is that in order to reach the extremely low levels of e.g. oxygen (together with the other extreme values of light elements in Fig.~\ref{fig:abundance}), hot-hydrogen burning is needed because it has the relevant side reactions (namely, the Mg-Al chain and the Ne-Na chain) that produce the observed light element ratios. But of course by burning hydrogen, helium is created. 


These problems indicate that our general understanding of the multiple population phenomenon is still far from perfect. For example, we may not have the complete picture of how the Mg-Al chain and the Ne-Na chain operate in massive and very massive stars. This would not be surprising, given the scarcity of observations when it comes to massive stars especially at low-metallicity \citep[e.g. Sect.~1 in][]{Kubatova:2018}.

Alternatively, there may be a way to separate Na, O, Mg and Al from He either inside the first generation of massive stars or in the second generation of low-mass stars (gravitational settling of elements may play some role?). That said, there may be a missing ingredient in our theory of the cluster gas dynamics \citep[e.g. the mixing of the pristine gas to the ejecta of massive stars may happen in an unexpected way; turbulence in particular, cf.][may play a role?]{Hopkins:2013} or even that of the star formation process itself. Also, the method of measuring He variations indirectly from photometry and isochrone fitting, although currently our most reliable method to infer He dispersions \citep{Cassisi:2017}, may in the future undergo some now not yet seen developments, leading to the revision of what we know about helium in GCs. 

And finally, although some observational features are well explained by the hypothesis that a first generation of massive stars synthesised the observed element ratios, the problem around He may still mean that the massive star hypothesis needs to be looked beyond, and completely new hypotheses need to be suggested \citep[see also the discussion in Sect.~2.2.2 of][]{Bastian:2018}. 

\subsection{Directions for future work}\label{sec:future}

We combined two research fields, massive stellar evolution and cluster gas dynamics; both have studied GCs so far mostly from their own perspectives. Combining them now opens up new pathways for investigation.

For example, we only applied single stellar models. But massive stars have a very high binary fraction \citep{Sana:2012}. In a close binary, the interaction may lead to rather interesting outcomes such as the formation of gravitational-wave progenitors \citep[e.g.][]{deMink:2009b,Belczynski:2016,deMink:2016,Marchant:2016,Marchant:2017,Szecsi:2017short,Szecsi:2017long,VignaGomez:2018}. In fact, interacting massive binaries have been suggested as the source of anomalous light element ratios by \citet{deMink:2009} and investigated further by \citet{Bastian:2013} and \citet{Elmegreen:2017}. We note that our single stellar models have been computed with very similar physical assumptions as the binary system in \citet{deMink:2009}. Therefore we expect our stellar models to, upon putting them next to a companion in such a system, pollute the cluster at some extent via the binary channel as well. This however needs to be quantitatively investigated in the future, preferably by computing a set of detailed binary models. This would be a long-awaited completion of the work by \citet{deMink:2009} who only computed one such system and extrapolated therefrom. Alternatively, although less elaborate, our set of single star models may be applied in a synthetic binary population, similarly to what we did here for a synthetic single star population, but with some necessary assumptions about the interaction of the companions. This binary population then can be applied as input for the cluster gas dynamics simulations, potentially allowing us to test uncertain parameters of stellar and binary evolution (e.g. mass transfer efficiency, nuclear reaction rates etc.).

Another way to make use of our new method, is to test the predictions of different pollution scenarios against each other. For example, the `fast spinning star' scenario \citep{Decressin:2007} can be tested against ours, or the interactive binary scenario against these etc., in terms of facilitating star formation and with which chemical composition. 
Alternatively, the role of rotation may be explored by using not only slow-rotating models of \citet{Szecsi:2015} but moderate, and even fast rotators that predict chemically homogeneously evolution. These so-called `TWUIN stars' \citep{Kubatova:2018} are needed for the star forming shell scenario of \citet{Szecsi:2018} to operate. So if we want to explain out our present work's insufficient accounting for the high extent of Mg depletion by the shell-scenario (as done in Sect.~\ref{sec:MgAl-lowZ}), we need to make sure that the contribution of hot TWUIN stars do not brake down our `window' of star formation by heating up the gas too much.

The role of metallicity should also be studied in more detail, as this may indeed be important especially when it comes to the most fragile elements (e.g. magnesium). \textbf{Also, the very massive supergiants, those responsible for accommodating the secondary starformation, are only predicted in our low-metallicity population and not in our high-metallicity one; but the exact metallicity value below which supergiants start to play their role in the formation of GCs, should be specified further in future work.} We suggest to apply several sets of stellar models covering the range in metallicities between, and even above and below, of the two values investigated here. 

\textbf{Our low-metallicity supergiants are theoretical predictions; if they indeed exist, their observational discovery is yet to be carried out. They are of course not expected to be found in globular clusters, because even if they used to be there, they are long dead for now. As our models show, they are only expected to be present in clusters as young as between $\sim$2$-$4~Myr. 
Another caveat for their discovery is that their natal cloud needs to be sufficiently metal-poor and sufficiently massive to be able to form them at all. Nonetheless, from the models we know that they may be extremely bright objects. \citet[][see their Sect.~5]{Szecsi:2015} explains that at the distance of 18~Mpc, they would appear with a visual magnitude of 19~mag. It has been suggested that brightness variations due to pulsations \citep[with periods of the order of months to years, see also][]{Moriya:2015} may reveal them as stars rather than star clusters in photometric multi-epoch observations.}

When it comes to the calculations of the cluster gas dynamics, there are ways to improve here too. We suggest for example to apply our stellar models (both the low- and high-metallicity ones) as input for 3D~radiation-hydrodynamic simulations. There are basically two ways to do this. One is to compute the so-called smooth source hydrodynamics of the cluster \citep[following][]{Wunsch:2017}, that is, to suppose a population of several hundred massive stars providing energy and mass which is inserted smoothly into the cluster. Another, more elaborate but also computationally more expensive way is to model the cluster evolution applying individual sources in the 3D simulations. 

Less concerned with stellar evolution or cluster gas dynamics, ways to improve our general understanding of the multiple population phenomenon has been pointed out throughout the text. To summarize these: the effect of supernovae on our star formation episode should be assessed (Sect.~\ref{sec:supernovae}), the conditions for dynamical mass removal \textbf{including its effect on changing the ratio of first vs. second generation stars} should be quantified  (Sect.~\ref{sec:dynamical}), and the extent of mixing the enriched material with the pristine gas in the cluster center should be investigated (Sect.~\ref{sec:NaO}). And last but not least, the issues around helium should be resolved (Sect.~\ref{sec:heliumproblem}).

\textbf{
An interesting observational conundrum that arose recently, is that there seems to be a cut in the occurrence of the multiple population phenomenon at $\sim$2~Gyr. Some clusters with this age, like NGC~1978, do show multiple populations, whereas slightly younger clusters with comparable present-day masses do not \citep{Mucciarelli:2008,Mucciarelli:2014,Martocchia:2018,Martocchia:2018b}. If this proves to be an established fact in the future, any theoretical models should be able to account for it. It will need to be done, however, by including the investigation of the cluster's longterm dynamical evolution. As we have no means to do that yet (see also our discussion in Sect.~\ref{sec:dynamical}), we leave this question open for now.}

\subsection{Gravitational-waves}

With direct detections of gravitational waves \citep{Abbott:2016b,Abbott:2016a,Bagoly:2016,Abbott:2017,Szecsi:2017short,Szecsi:2017long}, many authors suggested globular clusters to be the host of these events \citep{Rodriguez:2015,Antonini:2016,Belczynski:2016,Askar:2017}. 
In Sect.~\ref{sec:supernovae}, discussing the final fate and remnants of our supergiants, we pointed out that many of our stellar models are expected to form black holes or neutron stars after they explode; only those that explode as pair-instability supernova are not expected to leave any remnant. Thus, without quantitatively investigating this question, we point out that our scenario qualitatively predicts a significant number of these compact objects to be present in the young massive cluster, and thus to potentially merge over the long lifetimes of these clusters via dynamical interactions. Our work is therefore an important motivation to look for gravitational-wave emission, as well as its compact object progenitors, in globular clusters. The same is true for short-duration gamma-ray bursts, the origin of which have been associated with gravitational-wave emitting compact object mergers \citep{Abbott:2017a}. 

	
	\section{Conclusions}\label{sec:conclusions}
	
We realized a novel synergy between two research fields, massive stellar theory and cluster gas dynamics. In particular, we explored whether the model of rapidly cooling shocked stellar winds combined with state-of-the-art stellar evolution models can contribute to the explanation of multiple stellar populations observed in globular clusters.

The model of rapidly cooling shocked stellar winds predicts that the hot gas within star clusters can become thermally unstable and form warm clumps. These clumps fall into the cluster centre where they cool further and form a second generation of stars. The new stars are necessarily enriched by the nuclear ashes synthesised in the first generation massive stars.

We apply stellar evolutionary models as input for the calculations of the cluster gas dynamics. The models are computed for two chemical compositions: for low-metallicity corresponding to [Fe/H]~$\sim$~$-$1.7, and for a higher, but still subsolar metallicity corresponding to [Fe/H]~$\sim$~$-$0.4. By applying these two sets of models, we evaluate the impact of metallicity on the secondary star formation.

We find that at low-metallicity, cool supergiant stars---predicted to have very high mass loss rates and, in the same time, a low wind velocity---help to make the hot gas thermally unstable \textbf{very early on}. Their winds include products of hot hydrogen-burning, thus making them a suitable candidate for explaining the multiple population phenomenon. 

Our calculations are run for the initial 10~Myr life of the clusters, and predict the amount of mass accumulated inside the cluster center, as well as its chemical composition depending on the cluster mass, slope of the initial mass function and metallicity. We draw the following conclusions: 

\begin{enumerate}
\item \oldtextbf{A `window' for undisturbed star formation.} At low-metallicity, mass accummulation starts early (at $\sim$1.6~Myr), and a significant amount of mass is available for star formation before the first supernovae start to explode (at around 4.5~Myr) \textbf{or before the gas is expelled from young massive clusters (typically observed around 4~Myr)}. This is thanks to the slow winds of massive supergiants. At high-metallicity however, mass accumulation starts later, after $\sim$4~Myr, \textbf{thus limiting---but not necessarily excluding---our scenario to work.} 

\item \oldtextbf{Agreement with light-element abundance ratios.}
Our calculations reproduce the Na-O spread sufficiently well. Also, only hydrogen-burning products are ejected (i.e. the sum of the C, N and O atoms are preserved, with C \& O being depleted and N enhanced), but no products of helium-burning or those of subsequent burning stages. The spread our calculations predict in Mg-Al is lower than observed; although the stellar models do provide the right chemical composition (i.e. heavily depleted in Mg and enriched in Al), it is not captured by our `window' of star formation. This, together with predicting higher than observed helium abundances, points to future directions of improvement. 

\item \oldtextbf{Cluster center captures all the mass of stellar winds.}
Our metal-poor clusters with initial mass larger than several 10$^6$~M$_{\odot}$ capture almost all the mass ejected by stellar winds. This 10$^4$--10$^5$~M$_{\odot}$ material accumulates in the center of the cluster and forms new stars there.  Thus, under the assumption that the massive cluster is evolving to become a globular cluster by dynamically removing its not too centrally located (i.e. mainly first generation) stars over several gigayears, we are able to consistently explain the mass budget of present day globular clusters by applying a normal initial mass function for both the first, and the second, stellar generations.

\item \oldtextbf{A top-heavy IMF helps though.}
With a normal initial mass function ($\alpha_{4}$~$=$~$-$2.3) our massive clusters form a second generation as populous as 1\% of the total. But applying a top-heavy initial mass function with $\alpha_{4}$~$=$~$-$1.5 raises this fraction to close to 50\%, and a very extreme index of $\alpha_{4}$~$=$~$-$1.1 up to 80\%. This means that even without supposing dynamical removal of the first generation (see above), a top-heavy IMF can explain the observed high ratio of second vs. first generation of stars. If both effects contribute however, we suggest that a moderate amount of dynamical ejection together with a moderately top-heavy IMF is enough to account for the mass budget of present day globular clusters. 

\item \oldtextbf{More than two generations?} \textbf{In principle our scenario is able to produce more than two stellar generations since we allow the newly formed generation to contain massive stars, thus possibly repeating the cycle.} 

\item \oldtextbf{Globular clusters as hosts of gravitational wave emission.}
In our scenario, the second generation is formed from the \textit{winds} of massive stars. The massive stars themselves are predicted to end up mostly as compact objects, supporting the hypothesis that gravitational-waves should be expected from globular clusters. 

\item \oldtextbf{Fraction of second generation correlates with GC mass.} We predict that both the amount of accumulated mass and the total initial mass of the cluster correlate with the number of low-mass stars in the second generation. This may lead to the observed correlation between the mass of the GC and the extent of the multiple population phenomenon. 


\end{enumerate}

\ \ 

Our way of investigating the multiple population phenomenon in GCs by combining stellar evolutionary models with calculations of cluster wind dynamics, should be considered an important method of testing stellar theories in the future---especially those that are very hard to find observational evidence for. Such theories include metal-poor massive stars, both in single and binary systems. The potential of these systems in explaining exotic explosions (gravitational-waves, gamma-ray bursts, several types of supernovae and superluminous supernovae) is quite high; nonetheless, many of the existing theories awaits future tests. We suggest that studying GCs by combining stellar models with cluster wind dynamics is a viable new approach by which these tests could be done.

	
	\acknowledgments
	\oldtextbf{Acknowledgments.}
This work has been supported by the institutional project RVO:67985815. D.Sz.\ was supported by the Czech Science Foundation Grant nr.\ 16-01116S (GA \v{C}R). 
The authors offer special thanks to E.~Pancino for her kind help in providing us with the \textit{Gaia}/ESO data and for various related discussions. We are also thankful for the constructive comments of our anonymous referee, as well as to J.~Palou\v{s}, S. Ehlerov\'a, R. Taylor, S. Mart\'inez-Gonz\'alez, J.~Mackey, N.~Langer, G.~Gr\"afener and E.~Carretta for discussions on the subject. D.Sz. says further thanks to \'{A}.~Szab\'{o} for combing her hair when she really needed it. This research was partially supported by STFC.

	\software{Numpy}
	
	\bibliographystyle{yahapj}
	\bibliography{References}
	
	
\end{document}